\documentclass[aps,graphicx,12pt,showkeys,showpacs,onecolumn]{revtex4}
\usepackage{amsmath}
\usepackage{amscd}
\usepackage{graphicx}
\usepackage{subfigure}
\usepackage{appendix}

\begin{document}

\title[Short Title]{Fast and noise-resistant implementation of quantum phase gates and creation of quantum entangled states}

\author{Ye-Hong Chen$^{1}$}
\author{Yan Xia$^{1,}$\footnote{E-mail: xia-208@163.com}}
\author{Qing-Qin Chen$^{2}$}
\author{Jie Song$^{3,}$}

\affiliation{$^{1}$Department of Physics, Fuzhou University, Fuzhou
350002, China\\$^{2}$Zhicheng College, Fuzhou University, Fuzhou
350002, China\\$^{3}$Department of Physics, Harbin Institute of
Technology, Harbin 150001, China}

%\tableofcontents

\begin{abstract}
The ``Lewis-Riesenfeld phases'' which plays a crucial role in
constructing shortcuts to adiabaticity may be a resource for the
implementation of quantum phase gates. By combining
``Lewis-Riesenfeld invariants'' with ``quantum Zeno dynamics'', we
propose an effective scheme of rapidly implementing $\pi$ phase
gates via constructing shortcuts to adiabatic passage in a
two-distant-atom-cavity system. The influence of various decoherence
processes such as spontaneous emission and photon loss on the
fidelity is discussed. It is noted that this scheme is insensitive
to both error sources. Additionally, a creation of $N$-atom cluster
states is put forward as a typical example of the applications of
the fast and noise-resistant phase gates. The study results show
that the shortcuts idea is not only applicable in other logic gates
with different systems, but also propagable for many quantum
information tasks.
\end{abstract}


\maketitle

%\scriptsize

\section{Introduction}
The recent development of quantum information processing (QIP) has
shed new light on complexity and communication theory. The existing
quantum algorithms for specific problems shows that quantum
computers can in principle solve computational problems much more
efficiently than classical computers
\cite{DDPRSA85,LKGPrl97,PWSIEEE94,JLMPMSKSBPrl06,KBHTBWSRmp98,PKITMSRmp07,NVVTHBWSKBArpc01}.
 So far, a lot of studies on the theoretical and practical aspects of
quantum computation have been triggered in recent years
\cite{zjdengpra07,lmduanpra05}.  It has been shown that the key
ingredients of achieving quantum computation are quantum logic
gates. Using one-qubit and two-qubit logic gates, a universal
quantum computation network will be constructed. That is, any
unitary transformation, including multiqubit gates, can be
decomposed into a series of single qubit operations along with
two-qubit logic gates \cite{TSHWPrl95,DPDPra98}  in principle.
Therefore, this concept has intensively motivated the researches in
the realization of the one-qubit unitary gates and the two-qubit
logic gates in many physical systems
\cite{zjdengpra07,lmduanpra05,TSHWPrl95,DPDPra98,turchette,monroe,jones,ljsham,yamamoto,vanderwal}.
Among them, the cavity quantum electrodynamics (QED) systems which
open up new prospects in the implementation of large-scale quantum
computation and generation of nonclassical states with the atoms
trapped in optical cavities, have recently become very promising and
highly inventive for QIP
\cite{JMRMBSHRmp01,MASetalNat07,JMetalNat07,HJKNat08}. And there
have been tremendous advances in experiments realizing cavity QED
systems during the past few years, for example, single cesium atoms
have been cooled and trapped in a small optical cavity in the
strong-coupling regime
\cite{JMetalPrl03,JYDWVHJKPrl99,KMBetalNat05}; Quantum computations
have been realized in cavity QED \cite{turchette,ecqed1}. On the
other hand, based on cavity QED systems, lots of theoretical schemes
\cite{SBZGCGPrl00,SBXPrl05,CPYSICSHPra04,DJJICPZPrl00,ESMFSPMPra01,ABGSAPra04,zxlygpra07}
have been proposed to implement two-qubit logic gates with trapped
atoms. For instance, Zheng implemented a $\pi$ phase gate through
the adiabatic evolution in 2005 \cite{SBXPrl05}. The procedure of
acquiring the quantum phase gate is  simplified because the scheme's
system was not required to undergo a desired cyclic evolution. And
the fidelity for the scheme was said to be improved due to the
avoidance of the errors. It was such a novel paper that it aroused a
great deal of interest in using adiabatic evolution to implement
phase gates \cite{SBZPra04,SBZPra05,SJYXHSSJLGJNEpl07,DMLBMKMPra07}.

However, the adiabatic condition limits the speed of the evolution
of the system. The previous methods based on the adiabatic passage
technique usually require a relatively long interaction time. In
general, the interaction time for a method is the shorter the
better, otherwise, the method may be useless because the dissipation
caused by decoherence, noise, and losses on the target state
increases with the increasing of the interaction time. Therefore,
finding shortcuts to adiabaticity is of great significance for QIP
and has drawn much more attention than ever before. In recent years,
various reliable, fast, and robust schemes have been proposed in
finding shortcuts to a slow adiabatic passage in both theory
\cite{XCILARDGJGMPrl10,ETSLSMGMMADCDGOARXCJGMAmop13,ARXCDAJGMNjp12,AdCPra11,XCETJGMPra11,XCJGMPra12,MDSARJpca03Jpcb05Jcp08,MVBJpamt09}
and experiment
\cite{JFSXLSPVGLPra10,JFSXLSPCPVGLEpl11,AWFZTRSTDKOMHKSFSKUPPrl12,RBJGYLTRTDHJDJJPHDLDJWPrl12,SYTXCOl12,MGBMVNMPHEADCRFVGRMOMNat12}.
For example, Chen and Muga constructed shortcuts by invariant-based
inverse engineering to perform fast population transfer in
three-level systems in 2012 \cite{XCJGMPra12}. Lu \emph{et al.}
constructed shortcuts by using the approach of ``transitionless
quantum driving'' to the population transfer and the creation of
maximal entanglement in the cavity QED system
\cite{MLYXLTSJSNBAPra14}. In 2014, motivated by the quantum Zeno
dynamics \cite{PKHWTHAZMAKPrl95,PFSPPrl02}, Chen \emph{et al.}
constructed shortcuts to perform the fast populations transfer of
ground states in multiparticle systems with the invariant-based
inverse engineering \cite{YHCYXQQCJSPRA14}. Soon after that, they
generalized their method into distant cavities to realize fast and
noise-resistant QIP, including quantum population transfer,
entangled states¡¯ preparation and transition
\cite{YHCYXQQCJSLPL14}.

Refs. \cite{XCILARDGJGMPrl10,ETSLSMGMMADCDGOARXCJGMAmop13,XCJGMPra12,MVBJpamt09,MDSARJpca03Jpcb05Jcp08,YHCYXQQCJSPRA14}
prove that constructing shortcuts always has to deal with the
quantum phases transition, for example, the Berry phases
\cite{MVBJpamt09}, the Lewis-Riesenfeld (LR) \cite{HRLWBRJmp69}
phases, and so on. In view of that we wonder if it is possible to
use shortcuts to adiabatic passage (STAP) for the implementation of
logic gates. Therefore, in this paper, we effectively construct STAP
in a two-distant-atom system to implement $\pi$ phase gates
through designing resonant laser pulses by the invariant-based
inverse engineering. To the best of our knowledge, this is the first
scheme for quantum logic gates based on STAP in cavity QED systems.
The logic gates can be achieved in a much shorter interaction time
via STAP than that based on adiabatic passage and the interaction
time also not needs to be controlled accurately. Similar to the
adiabatic passage, the present scheme is insensitive to variations
of the parameters. Moreover, because of the short interaction time,
the present scheme is insensitive to the decoherence caused by
spontaneous emission and photon loss even though we increase the
populations of some excited states during the gate operation. And we
also give an example of applications of this kind of phase gates:
using the fast and noise-resistant gates to create $N$-atom cluster
states in spatially-separated cavities.

This paper is structured as follows. In Sec. \textrm{II}, we briefly
describe the LR phases. In Sec. \textrm{III}, we construct STAP in a
two-distant-atom system by the invariant-based inverse
engineering, and show how to use the shortcut to implement $\pi$
phase gates. Then, in Sec. \textrm{IV} we use the kind of phase
gates to create $N$-atom cluster states in $N$ distant cavities
connected by $N-1$ fibers. We give the numerical simulation and
experimental discussion about the validity of the scheme in Sec.
\textrm{V}. And Sec. \textrm{VI} is the conclusion.

\section{Lewis-Riesenfeld phases}
A brief description about the LR theory \cite{HRLWBRJmp69} is given in this section.
We consider a time-dependent quantum system whose Hamiltonian is $H(t)$. Associated with the Hamiltonian, there are time-dependent Hermitian invariants of motion $I(t)$ that satisfy
\begin{eqnarray}\label{eq2-1}
  i\hbar {\partial_{t} I(t)}-[H(t),I(t)]=0,
\end{eqnarray}
where $\partial_{t}\equiv\frac{\partial}{\partial t}$. For any solution $|\Psi(t)\rangle$ of
the time-dependent Schr\"{o}dinger equation $i\hbar \partial_{t}|\Psi(t)\rangle=H(t)|\Psi(t)\rangle$,
$I|\Psi(t)\rangle$ is also a solution \cite{MALJpa09}, and $|\Psi(t)\rangle$ can be expressed as a linear combination of invariant modes
\begin{eqnarray}\label{eq2-2}
  |\Psi(t)\rangle=\sum_{n}C_{n}e^{i\alpha_{n}}|\phi_{n}(t)\rangle,
\end{eqnarray}
where $C_{n}$ are constants, $|\phi_{n}(t)\rangle$ is the $n$th eigenvector of $I(t)$ and the corresponding real eigenvalue is $\lambda_{n}$.
The LR phases $\alpha_{n}$ fulfill
\begin{eqnarray}\label{eq2-3}
  \hbar\frac{d\alpha_{n}}{dt}=\langle\phi_{n}(t)|i\hbar\partial_{t}-H(t)|\phi_{n}(t)\rangle.
\end{eqnarray}

\section{Shortcut to Adiabaticity for two-qubit $\pi$ phase gates.}

\begin{figure}
 \scalebox{0.35}{\includegraphics {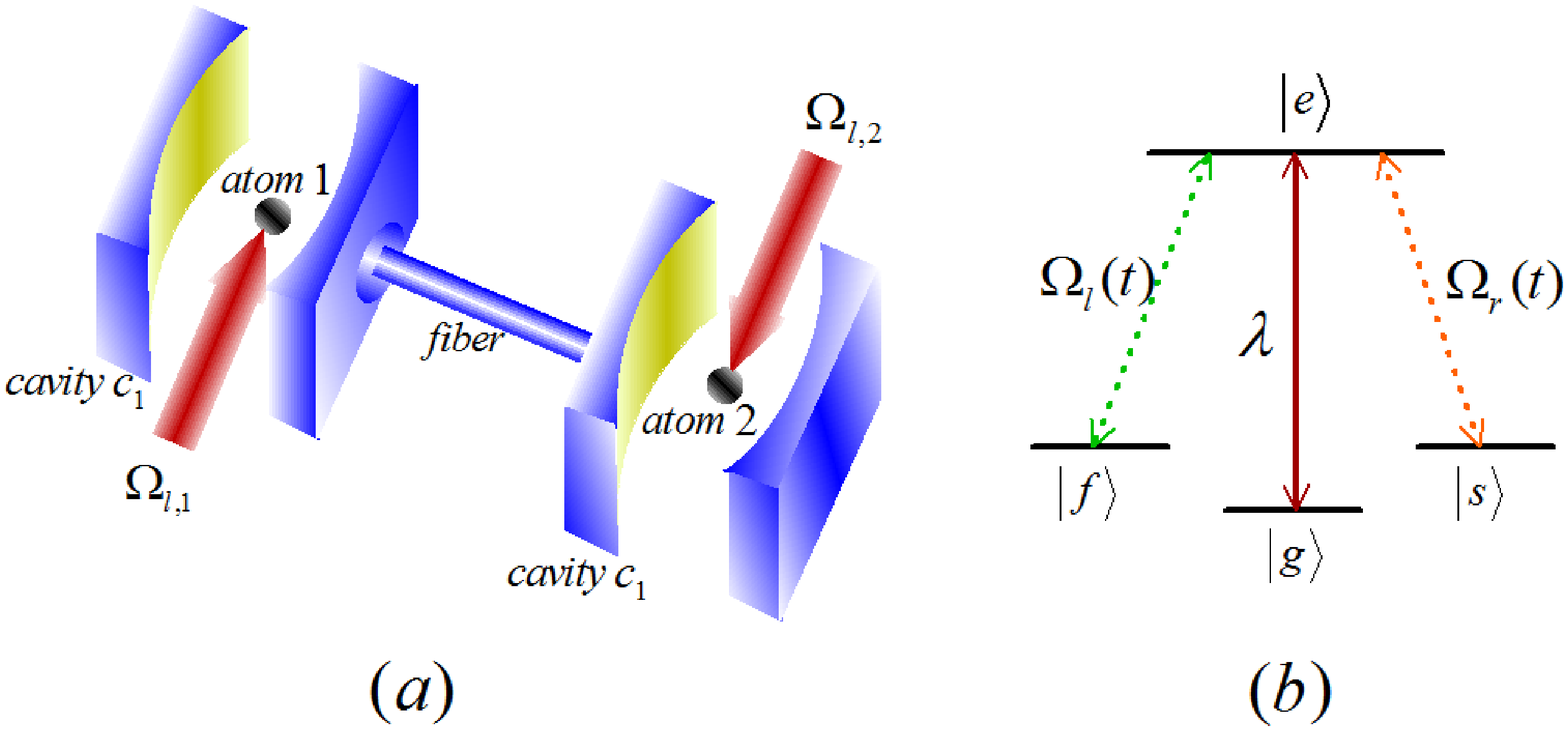}}
 \caption{The cavity-fiber-atom combined system and the atomic level configuration for the implementation of phase gates.}
 \label{model}
\end{figure}

We consider that two four-level atoms (1 and 2) are respectively
trapped in two cavities ($c_1$ and $c_2$) connected by a fiber as
shown in Fig. \ref{model}. Each atom has three ground states
$|g\rangle$, $|f\rangle$ and $|s\rangle$, and an excited state
$|e\rangle$. The whole system evolves in the one-excited subspace
spanned by
\begin{eqnarray}\label{eq3-0}
  |\psi_{1}\rangle&=&|f\rangle_{1}|g\rangle_{2}|0\rangle_{c_{1}}|0\rangle_{c_{2}}|0\rangle_{f}, \cr
  |\psi_{2}\rangle&=&|e\rangle_{1}|g\rangle_{2}|0\rangle_{c_{1}}|0\rangle_{c_{2}}|0\rangle_{f}, \cr
  |\psi_{3}\rangle&=&|g\rangle_{1}|g\rangle_{2}|1\rangle_{c_{1}}|0\rangle_{c_{2}}|0\rangle_{f}, \cr
  |\psi_{4}\rangle&=&|g\rangle_{1}|g\rangle_{2}|0\rangle_{c_{1}}|0\rangle_{c_{2}}|1\rangle_{f}, \cr
  |\psi_{5}\rangle&=&|g\rangle_{1}|g\rangle_{2}|0\rangle_{c_{1}}|1\rangle_{c_{2}}|0\rangle_{f}, \cr
  |\psi_{6}\rangle&=&|g\rangle_{1}|e\rangle_{2}|0\rangle_{c_{1}}|0\rangle_{c_{2}}|0\rangle_{f}, \cr
  |\psi_{7}\rangle&=&|g\rangle_{1}|f\rangle_{2}|0\rangle_{c_{1}}|0\rangle_{c_{2}}|0\rangle_{f}, \cr
  |\psi_{8}\rangle&=&|s\rangle_{1}|g\rangle_{2}|0\rangle_{c_{1}}|0\rangle_{c_{2}}|0\rangle_{f}, \cr
  |\psi_{9}\rangle&=&|g\rangle_{1}|s\rangle_{2}|0\rangle_{c_{1}}|0\rangle_{c_{2}}|0\rangle_{f}.
\end{eqnarray}
We assume that the atomic transition
$|f\rangle\leftrightarrow|e\rangle$ is resonantly driven through
a time-dependent laser pulse with Rabi frequency
$\Omega_{l}(t)$, transition $|s\rangle\leftrightarrow|e\rangle$ is resonantly driven through
a time-dependent laser pulse with Rabi frequency
$\Omega_{r}(t)$, the transition $|g\rangle\leftrightarrow|e\rangle$ is resonantly coupled to the cavity mode with coupling
constant $\lambda$. In the short fiber limit $L\tau/(2\pi c)\ll1$ \cite{PTPrl97,SAMSBSPrl06},
where $L$ denotes the fiber length, $c$ denotes the speed
of light, and $\tau$ denotes the decay of the cavity field into a continuum
of fiber mode, the cavities $c_{1}$ and $c_{2}$ are
coupled to one fiber mode with strength $v$. We turn off the classical laser fields $\Omega_{r,k}$ ($k=1,2$) at first.
The Hamiltonian of the system in interaction picture for this case can be written as ($\hbar=1$)
\begin{eqnarray}\label{eq3-1}
  H_{i}(t)&=&H_{al}+H_{ac}+H_{cf},                                            \cr\cr
  H_{al}(t)&=&\sum_{k=1,2}{\Omega_{l,k}(t)|e\rangle_{k}\langle f|}+H.c.,        \cr\cr
  H_{ac}(t)&=&\sum_{k=1,2}{\lambda_{k}a_{k}|e\rangle_{k}\langle g|}+H.c.,           \cr\cr
  H_{cf}(t)&=&vb^{\dag}(a_{1}+a_{2})+H.c.,
\end{eqnarray}
where subscript $k$ represents the $k$th atom and $a_{k}$ is the annihilation operator for the $k$th cavity mode, and
$b^{\dag}$ represents the creation operator of the fiber mode.
Considering quantum Zeno dynamics, we set $\Omega_{l,k}(t)\ll \lambda$ (the Zeno condition)
and rewrite the Hamiltonian in eq. (\ref{eq3-1}) with the eigenvectors
of Hamiltonian $H_{im}$ ($H_{im}=H_{ac}+H_{cf}$) as ($g_{1}=g_{2}=g$ and $\chi=\sqrt{2v^{2}+\lambda^{2}}$)
\begin{eqnarray}\label{eq3-2}
  H_{I}&=&\sum_{m=0}^{4}\xi_{m}|\theta_{m}\rangle\langle\theta_{m}|
                     +[\frac{v}{\chi}|\theta_{0}\rangle(\Omega_{l,1}\langle\psi_{1}|+\Omega_{l,2}\langle\psi_{7}|)                           \cr\cr
                   &&+\frac{1}{2}(|\theta_{1}\rangle+|\theta_{2}\rangle)(-\Omega_{l,1}\langle\psi_{1}|+\Omega_{l,2}\langle\psi_{7}|)        \cr\cr
                   &&+\frac{\lambda}{2\chi}(|\theta_{3}\rangle+|\theta_{4}\rangle)(\Omega_{l,1}\langle\psi_{1}|+\Omega_{l,2}\langle\psi_{7}|)+H.c.],
\end{eqnarray}
where
\begin{eqnarray}\label{eq3-3}
  |\theta_{0}\rangle&=&\frac{v}{\chi}(|\psi_{2}\rangle-\frac{\lambda}{v}|\psi_{4}\rangle+|\psi_{6}\rangle),                         \cr
  |\theta_{1}\rangle&=&\frac{1}{2}(-|\psi_{2}\rangle-|\psi_{3}\rangle+|\psi_{5}\rangle+|\psi_{6}\rangle),                           \cr
  |\theta_{2}\rangle&=&\frac{1}{2}(-|\psi_{2}\rangle+|\psi_{3}\rangle-|\psi_{5}\rangle+|\psi_{6}\rangle),                           \cr
  |\theta_{3}\rangle&=&\frac{\lambda}{2\chi}
                     (|\psi_{2}\rangle
                     +\frac{\chi}{\lambda}|\psi_{3}\rangle
                     +\frac{2v}{\lambda}|\psi_{4}\rangle
                     +\frac{\chi}{\lambda}|\psi_{5}\rangle
                     +|\psi_{6}\rangle),                                                                                            \cr
  |\theta_{4}\rangle&=&\frac{\lambda}{2\chi}
                     (|\psi_{2}\rangle
                     -\frac{\chi}{\lambda}|\psi_{3}\rangle
                     +\frac{2v}{\lambda}|\psi_{4}\rangle
                     -\frac{\chi}{\lambda}|\psi_{5}\rangle
                     +|\psi_{6}\rangle),
\end{eqnarray}
are the eigenvectors of $H_{im}$ corresponding eigenvalues
$\xi_{0}=0$, $\xi_{1}=\lambda$, $\xi_{2}=-\lambda$, $\xi_{3}=\chi$ and $\xi_{4}=-\chi$, respectively.
Perform the unitary transformation $U=e^{-i(\sum_{m=0}^{4}{\xi_{m}|\theta_{m}\rangle\langle\theta_{m}|})t}$, and we obtain
\begin{eqnarray}\label{eq3-4}
  H_{app}&=&\frac{1}{2}(e^{i\lambda t}|\theta_{1}\rangle+e^{-i\lambda t}|\theta_{2}\rangle)(-\Omega_{l,1}\langle\psi_{1}|+\Omega_{l,2}\langle\psi_{7}|)              \cr\cr
                   &&+\frac{\lambda}{2\chi}(e^{i\chi t}|\theta_{3}\rangle+e^{-i\chi t}|\theta_{4}\rangle)(\Omega_{l,1}\langle\psi_{1}|+\Omega_{l,2}\langle\psi_{7}|)  \cr\cr
                   &&+\frac{v}{\chi}|\theta_{0}\rangle(\Omega_{l,1}\langle\psi_{1}|+\Omega_{l,2}\langle\psi_{7}|)  +H.c..
\end{eqnarray}
Neglecting the terms with high oscillating frequencies $\lambda$ and $\chi$ then the final effective Hamiltonian is gotten,
\begin{eqnarray}\label{eq3a-5}
  H_{eff}&=&\frac{v}{\chi}|\theta_{0}\rangle(\Omega_{l,1}\langle\psi_{1}|+\Omega_{l,2}\langle\psi_{7}|)+H.c. \cr
         &=&\frac{v}{\chi}
              \left(
                \begin{array}{ccc}
                 0 & \Omega_{l,1}(t) & 0                \\
                 \Omega_{l,1}(t) & 0 & \Omega_{l,2}(t)  \\
                 0 & \Omega_{l,2}(t) & 0                \\
              \end{array}
             \right).
\end{eqnarray}
It is obvious that $H_{eff}$ possesses SU(2) dynamical symmetry \cite{YZLJQLHJWMKJGZPra96}, and the invariant $I(t)$ is given by \cite{XCETJGMPra11,YHCYXQQCJSPRA14}
\begin{eqnarray}\label{eq3a-6}
  I(t)&=&\frac{v}{\chi}\omega_{0}[|\theta_{0}\rangle(\cos{\gamma}\sin{\beta}\langle\psi_{1}|+\cos{\gamma}\cos{\beta}\langle\psi_{7}|)
          -i\sin{\gamma}|\psi_{7}\rangle\langle\psi_{1}|]+H.c. \cr
      &=&\frac{v}{\chi}\omega_{0}
              \left(
                \begin{array}{ccc}
                 0 & \cos{\gamma}\sin{\beta} & -i\sin{\gamma}           \\
                 \cos{\gamma}\sin{\beta} & 0 & \cos{\gamma}\cos{\beta}  \\
                 i\sin{\gamma} & \cos{\gamma}\cos{\beta} & 0            \\
              \end{array}
             \right).
\end{eqnarray}
where $\omega_{0}$ is an arbitrary constant with units of frequency to keep $I(t)$ with dimensions of energy.
By solving the differential equation in eq. (\ref{eq2-1}), the time-dependent Rabi frequencies $\Omega_{l,1}(t)$ and $\Omega_{l,2}(t)$
are given,
\begin{eqnarray}\label{eq3a-7}
  \Omega_{l,1}(t)&=&\frac{\chi}{v}(\dot{\beta}\cot{\gamma}\sin{\beta}+\dot{\gamma}\cos{\beta}), \cr
  \Omega_{l,2}(t)&=&\frac{\chi}{v}(\dot{\beta}\cot{\gamma}\cos{\beta}-\dot{\gamma}\sin{\beta}).
\end{eqnarray}
And the eigenstates of $I(t)$, with eigenvalues $\lambda_{0}=0$ and $\lambda_{\pm}=\pm1$, are
\begin{eqnarray}\label{eq3a-8}
  |\phi_{0}(t)\rangle&=&
    \left(
     \begin{array}{c}
       \cos{\gamma}\cos{\beta}      \\
       -i\sin{\gamma}               \\
       -\cos{\gamma}\sin{\beta}     \\
     \end{array}
    \right),   \cr\cr
  |\phi_{\pm}(t)\rangle&=&
    \frac{1}{\sqrt{2}}
    \left(
     \begin{array}{c}
       \sin{\gamma}\cos{\beta}\pm i\sin{\beta}    \\
       i\cos{\gamma}                              \\
       -\sin{\gamma}\sin{\beta}\pm i\cos{\beta}   \\
     \end{array}
    \right).
\end{eqnarray}
In order to achieve fast adiabatic-like process (a process which is not really adiabatic but leading to the same final populations),
we impose relations $[I(0),H_{eff}(0)]=0$ and $[I(t_{f}),H_{eff}(t_{f})]=0$, where $t_{f}$ is the total interaction time.
Thus, we obtain the boundary conditions ${\dot{\gamma}(0)}={\dot{\gamma}(t_{f})}=0$, and
the LR phases
\begin{eqnarray}\label{eq3-9}
  \alpha_{0}=0,
  \alpha_{\pm}=\mp\int_{0}^{t_{f}}[\dot{\beta}\sin{\gamma}+\frac{v}{\chi}(\Omega_{l,1}\sin{\beta}+\Omega_{l,2}\cos{\beta})\cos{\gamma}]dt.
\end{eqnarray}
Moreover, the constants $C_{n}$ ($n=0,\pm$) also can be given through the relation $C_{n}=\langle\phi_{n}(0)|\psi_{1}\rangle$.
Generating a $\pi$ phase change of the quantum states involved in such a system, the final state
should become $-|\psi_{1}\rangle$. Therefore, we choose the feasible parameters $\gamma(t)=\epsilon$ and $\beta(t)=\pi t/t_{f}$,
where $\epsilon$ is a small value. We obtain
\begin{eqnarray}\label{eq3-10}
  \Omega_{l,1}(t)&=&\Omega_{0}\sin{\beta}, \cr
  \Omega_{l,2}(t)&=&\Omega_{0}\cos{\beta},
\end{eqnarray}
and the final state
\begin{eqnarray}\label{eq3-11}
  |\Psi(t_{f})\rangle=
    \left(
     \begin{array}{c}
       -\cos^{2}{\epsilon}-\cos{\alpha}\sin^{2}{\epsilon}                                \\
       -i\sin{\epsilon}\cos{\epsilon}+i\cos{\alpha}\sin{\epsilon}\cos{\epsilon}          \\
       \sin{\epsilon}\sin{\alpha}                                                                      \\
     \end{array}
    \right),
\end{eqnarray}
where $\alpha=\pi/({2\sin{\epsilon}})$ and $\Omega_{0}=\chi\pi\cot{\epsilon}/(vt_{f})$. When we choose $(\sin{\epsilon})^{-1}=4\zeta$ ($\zeta=1,2,3,\cdots$),
$\cos{\alpha}=1$ and the final state becomes $|\Psi(t_{f})\rangle=[-1,0,0]'=-|\psi_{1}\rangle$.

If the initial state of the system is $|f\rangle_{1}|s\rangle_{2}|0\rangle_{c}$, the whole
system evolves in a subspace spanned by
\begin{eqnarray}\label{eq3-12}
  |\varphi_1\rangle&=&|f\rangle_{1}|s\rangle_{2}|0\rangle_{c_{1}}|0\rangle_{c_{2}}|0\rangle_{f},   \cr
  |\varphi_2\rangle&=&|e\rangle_{1}|s\rangle_{2}|0\rangle_{c_{1}}|0\rangle_{c_{2}}|0\rangle_{f},   \cr
  |\varphi_3\rangle&=&|g\rangle_{1}|s\rangle_{2}|1\rangle_{c_{1}}|0\rangle_{c_{2}}|0\rangle_{f},   \cr
  |\varphi_4\rangle&=&|g\rangle_{1}|s\rangle_{2}|0\rangle_{c_{1}}|0\rangle_{c_{2}}|1\rangle_{f},   \cr
  |\varphi_5\rangle&=&|g\rangle_{1}|s\rangle_{2}|0\rangle_{c_{1}}|1\rangle_{c_{2}}|0\rangle_{f},
\end{eqnarray}
and the interaction picture Hamiltonian is
\begin{eqnarray}\label{eq3-13a}
  H_{i,s}=\Omega_{l,1}|\varphi_{2}\rangle\langle \varphi_{1}|+\lambda|\varphi_{2}\rangle\langle\varphi_{3}|+v|\varphi_{4}\rangle(\langle\varphi_{3}|+\langle\varphi_{5}|)+H.c..
\end{eqnarray}
With the parameters above, the adiabatic condition
$\langle\tilde{\theta}_{0}|\partial_{t}\tilde{\theta}_{\pm}\rangle\ll \tilde{\xi_{\pm}}$
for this subspace is fulfilled, where $|\tilde{\theta}_{0}\rangle$ is the eigenstate corresponding eigenvalue $\tilde{\xi}_{0}=0$,
and $|\tilde{\theta}_{\pm}\rangle$ are the eigenstates corresponding nonzero eigenvalues $\tilde{\xi}_{\pm}$. And if the initial state of system
is $|\tilde{\theta}_{0}(0)\rangle$, it evolves along the dark state
\begin{eqnarray}\label{eq3-13}
  |\tilde{\theta}_{0}(t)\rangle=\frac{1}{\sqrt{2\Omega_{l,1}^{2}+\lambda^{2}}}(\lambda|\varphi_{1}\rangle-\Omega_{l,1}(t)|\varphi_{3}\rangle+\Omega_{l,1}(t)|\varphi_{5}\rangle).
\end{eqnarray}
When $t=t_{f}$, $\Omega_{l,1}(t_{f})=0$ and the final state becomes $|\tilde{\theta}_{0}(t_{f})\rangle=|\varphi_{1}\rangle$.

It is obvious that the total Hamiltonian $H_{i}$ has no effect on the evolution of initial states
$|g\rangle_{1}|s\rangle_{2}|0\rangle_{c_{1}}|0\rangle_{c_{2}}|0\rangle_{f}$
and $|g\rangle_{1}|g\rangle_{2}|0\rangle_{c_{1}}|0\rangle_{c_{2}}|0\rangle_{f}$,
\begin{eqnarray}\label{eq3-14}
  H_{i}|g\rangle_{1}|s\rangle_{2}|0\rangle_{c_{1}}|0\rangle_{c_{2}}|0\rangle_{f}&=&0, \cr
  H_{i}|g\rangle_{1}|g\rangle_{2}|0\rangle_{c_{1}}|0\rangle_{c_{2}}|0\rangle_{f}&=&0.
\end{eqnarray}
Therefore, we have
\begin{eqnarray}\label{eq3-15}
  |g,s\rangle_{1,2}|0,0,0\rangle_{c_{1},c_{2},f}&\rightarrow&|g,s\rangle_{1,2}|0,0,0\rangle_{c_{1},c_{2},f}, \cr
  |g,g\rangle_{1,2}|0,0,0\rangle_{c_{1},c_{2},f}&\rightarrow&|g,g\rangle_{1,2}|0,0,0\rangle_{c_{1},c_{2},f}, \cr
  |f,s\rangle_{1,2}|0,0,0\rangle_{c_{1},c_{2},f}&\rightarrow&|f,s\rangle_{1,2}|0,0,0\rangle_{c_{1},c_{2},f}, \cr
  |f,g\rangle_{1,2}|0,0,0\rangle_{c_{1},c_{2},f}&\rightarrow&-|f,g\rangle_{1,2}|0,0,0\rangle_{c_{1},c_{2},f},
\end{eqnarray}
which is a $\pi$ phase gate.

\section{Preparation of cluster states}

\begin{figure}
 \scalebox{0.35}{\includegraphics {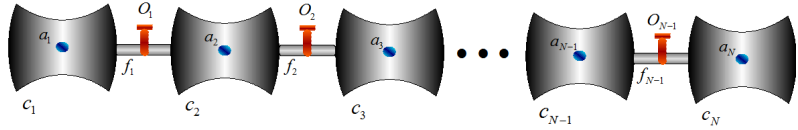}}
 \caption{The set-up diagram for the creation of cluster states.}
 \label{model2}
\end{figure}

There are some applications of the phase gates proposed in Sec. III,
say, it can be used to create cluster states. As shown in Fig.
\ref{model2}, $N$  atoms (1, 2, 3, $\cdots$, $N$) are trapped in $N$
cavities ($c_1$, $c_2$, $\ldots$ , $c_N$) connected by $N-1$ fibers
and optical switches \cite{JSYXHSSEpl09}, respectively. All the
optical switches ($O_{1}$, $O_{2}$, $\ldots$ , $O_{N-1}$) are turned
off in the beginning to void interaction with each atom. Then, we
turn on the optical switch $O_{1}$ to let the the optical fiber
mediate the cavities $c_{1}$ and $c_{2}$. And we prepare the atoms
$1$ and $2$ in the state
$|\varpi_{1}\rangle=\frac{1}{2}(|f\rangle_{1}+|g\rangle_{1})(|s\rangle_{2}+|g\rangle_{2})$.
Using the $\pi$ phase gate achieved above in Sec. III, we easily
obtain a two atoms state as follows:
\begin{eqnarray}\label{eq4-1}
  |\varpi_{1}'\rangle&=&\frac{1}{2}(|f,s\rangle_{1,2}-|f,g\rangle_{1,2}+|g,s\rangle_{1,2}+|g,g\rangle_{1,2})     \cr\cr
                     &=&\frac{1}{2}(|f\rangle_{1}\sigma^{z}_{2}+|g\rangle_{1})(|s\rangle_{2}+|g\rangle_{2}),
\end{eqnarray}
where $\sigma^{z}_{2}=|s\rangle_{2}\langle s|-|g\rangle_{2}\langle g|$. Then we turn off the optical switch $O_{1}$ and turn on $O_{2}$.
Meanwhile, the atom $3$ is prepared in state $\frac{1}{\sqrt{2}}(|f\rangle_{3}+|g\rangle_{3})$. In this case, the whole atomic system
is
\begin{eqnarray}\label{eq4-2}
  |\varpi_{2}\rangle=\frac{1}{2}(|f\rangle_{1}\sigma^{z}_{2}+|g\rangle_{1})(|s\rangle_{2}+|g\rangle_{2})\otimes\frac{1}{\sqrt{2}}(|f\rangle_{3}+|g\rangle_{3}).
\end{eqnarray}
 Afterward, we turn off laser pulse $\Omega_{l,k}$ (here $k=2,3$) and turn on the laser pulses $\Omega_{r,k}$. It is obvious that the behavior of $\Omega_{r}$ are the same as that of $\Omega_{l}$. The Hamiltonian for the whole system can be expressed as
\begin{eqnarray}\label{eq4-3}
  H_{i,2}(t)=vb_{2}^{\dag}(a_{2}+a_{3})+\sum_{k=2,3}{\Omega_{r,k}(t)|e\rangle_{k}\langle s|
                          +\lambda_{k}a_{k}|e\rangle_{k}\langle g|}+H.c.,
\end{eqnarray}
which is similar to the Hamiltonian in eq. (\ref{eq3-1}). Therefore, by using the same method in section \textrm{III}
(setting $\Omega_{r,2}=\Omega_{l,1}$ and $\Omega_{r,3}=\Omega_{l,2}$), we
achieve another $\pi$ phase gate:
\begin{eqnarray}\label{eq4-4}
  |g,f\rangle_{2,3}|0,0,0\rangle_{c_{2},c_{3},f_{2}}&\rightarrow&|g,f\rangle_{2,3}|0,0,0\rangle_{c_{2},c_{3},f_{2}}, \cr
  |g,g\rangle_{2,3}|0,0,0\rangle_{c_{2},c_{3},f_{2}}&\rightarrow&|g,g\rangle_{2,3}|0,0,0\rangle_{c_{2},c_{3},f_{2}}, \cr
  |s,f\rangle_{2,3}|0,0,0\rangle_{c_{2},c_{3},f_{2}}&\rightarrow&|s,f\rangle_{2,3}|0,0,0\rangle_{c_{2},c_{3},f_{2}}, \cr
  |s,g\rangle_{2,3}|0,0,0\rangle_{c_{2},c_{3},f_{2}}&\rightarrow&-|s,g\rangle_{2,3}|0,0,0\rangle_{c_{2},c_{3},f_{2}}.
\end{eqnarray}
Using this phase gate, the state $|\varpi_{2}\rangle$ becomes a three-atom cluster state:
\begin{eqnarray}\label{eq4-5}
  |\varpi_{2}'\rangle&=&\frac{1}{\sqrt{2}}(|f\rangle_{1}\sigma^{z}_{2}+|g\rangle_{1})
                        \frac{1}{2}(|s,f\rangle_{2,3}-|s,g\rangle_{2,3}+|g,f\rangle_{2,3}+|g,g\rangle_{2,3})  \cr\cr
                     &=&\frac{1}{2\sqrt{2}}(|f\rangle_{1}\sigma^{z}_{2}+|g\rangle_{1})(|s\rangle_{2}\sigma^{z}_{3}+|g\rangle_{2})(|f\rangle_{3}+|g\rangle_{3}),
\end{eqnarray}
where $\sigma^{2}_{3}=|f\rangle_{3}\langle f|-|g\rangle_{3}\langle
g|$. The rest may be deduced by analogy. Once an $(M-1)$-atom
($2\leq M\leq N$) cluster state has been generated, an $M$-atom
cluster state can be easily created by the following four steps:

(1) Turn on the optical switch $O_{M-1}$ (other optical switches are all kept off).

(2) (a) If $M$ is an odd number, the atom $M$ should be prepared in state $\frac{1}{\sqrt{2}}(|f\rangle_{M}+|g\rangle_{M})$, the
laser pluses $\Omega_{r,M}$ and $\Omega_{r,M-1}$ should be turned on.
(b) If $M$ is an even number, the atom $M$ should be prepared in state $\frac{1}{\sqrt{2}}(|s\rangle_{M}+|g\rangle_{M})$, the
laser pluses $\Omega_{l,M}$ and $\Omega_{l,M-1}$ should be turned on.

(3) Using the same method as in section III to implement the $\pi$ phase gate operation.

(4) The  $M$-atom cluster state can be generated through above three
steps.

\section{Numerical simulation and discussion}

\begin{figure}
 \renewcommand\figurename{\small FIG.}
 \centering \vspace*{8pt} \setlength{\baselineskip}{10pt}
 \subfigure[]{
\includegraphics[scale = 0.2]{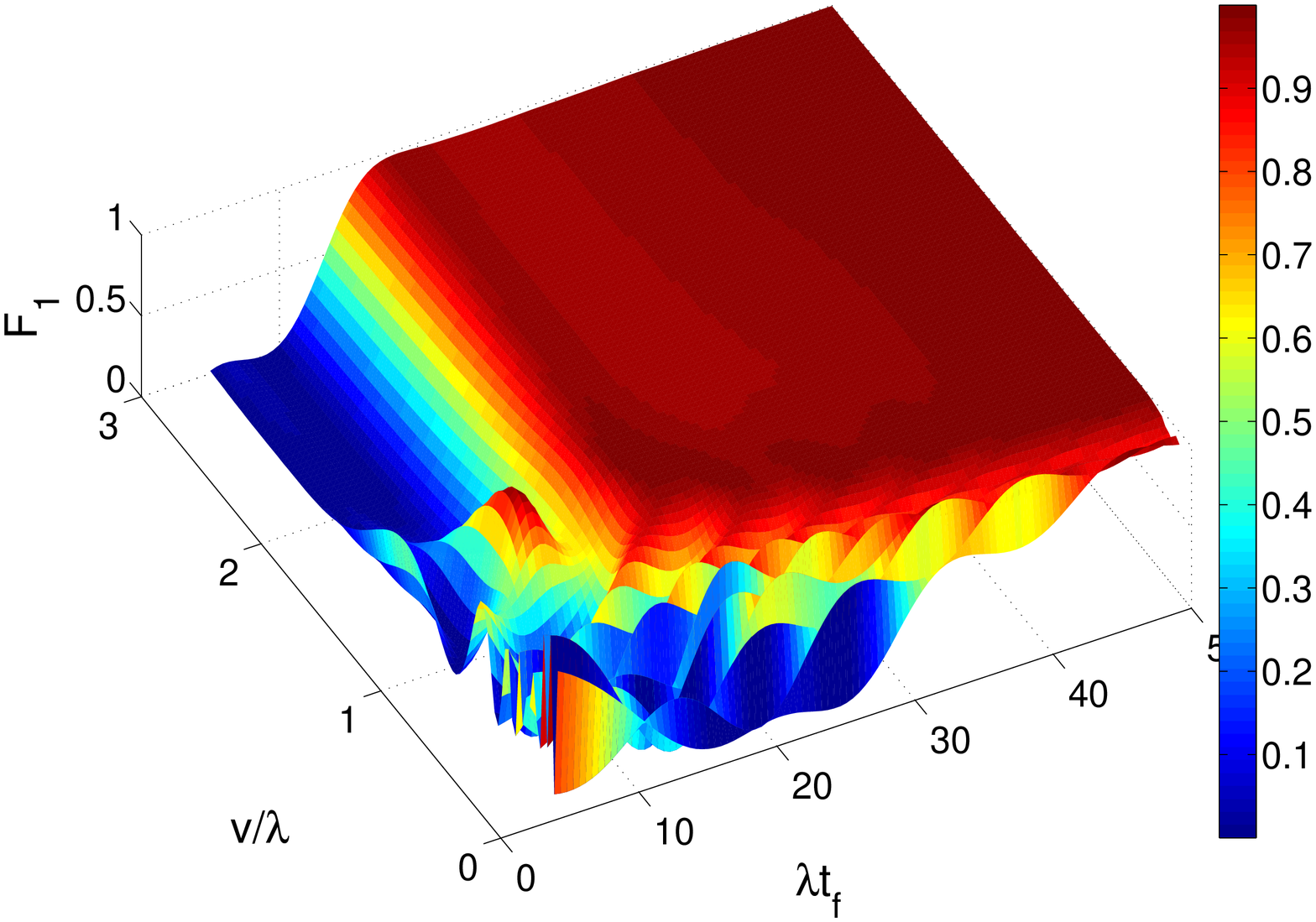}}
 \subfigure[]{
 \includegraphics[scale = 0.17]{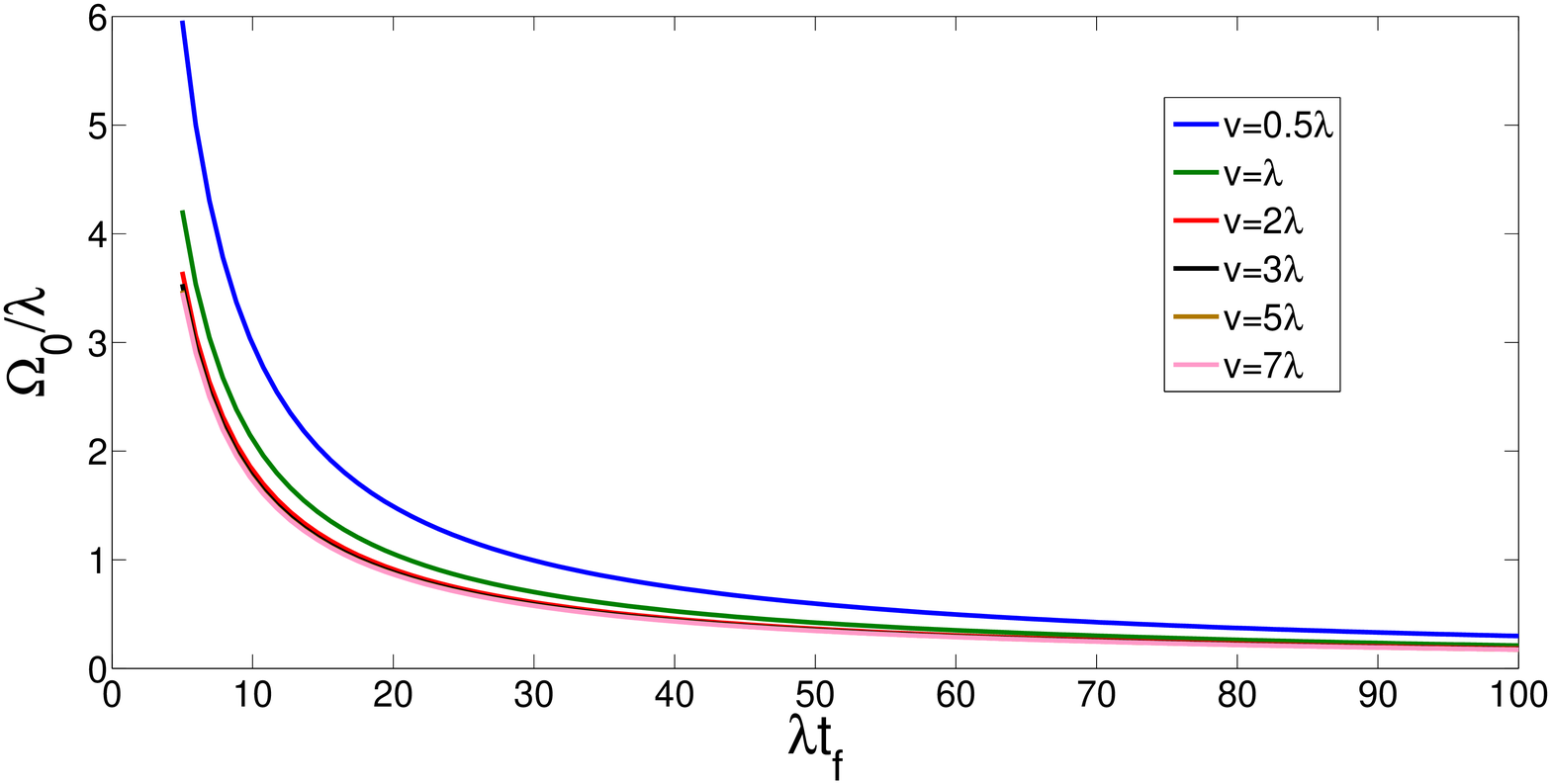}}
 \caption{
    (a) The fidelity $F_{1}$ of the target state $-|\psi_{1}\rangle$ versus the interaction time $\lambda t_{f}$ and the hopping rate $v/\lambda$.
    (b) The laser amplitude $\Omega_{0}/\lambda$ versus the interaction time $\lambda t_{f}$ with different hopping rates $v/\lambda$.
          }
 \label{Fvtf}
\end{figure}

\begin{figure}
 \scalebox{0.17}{\includegraphics {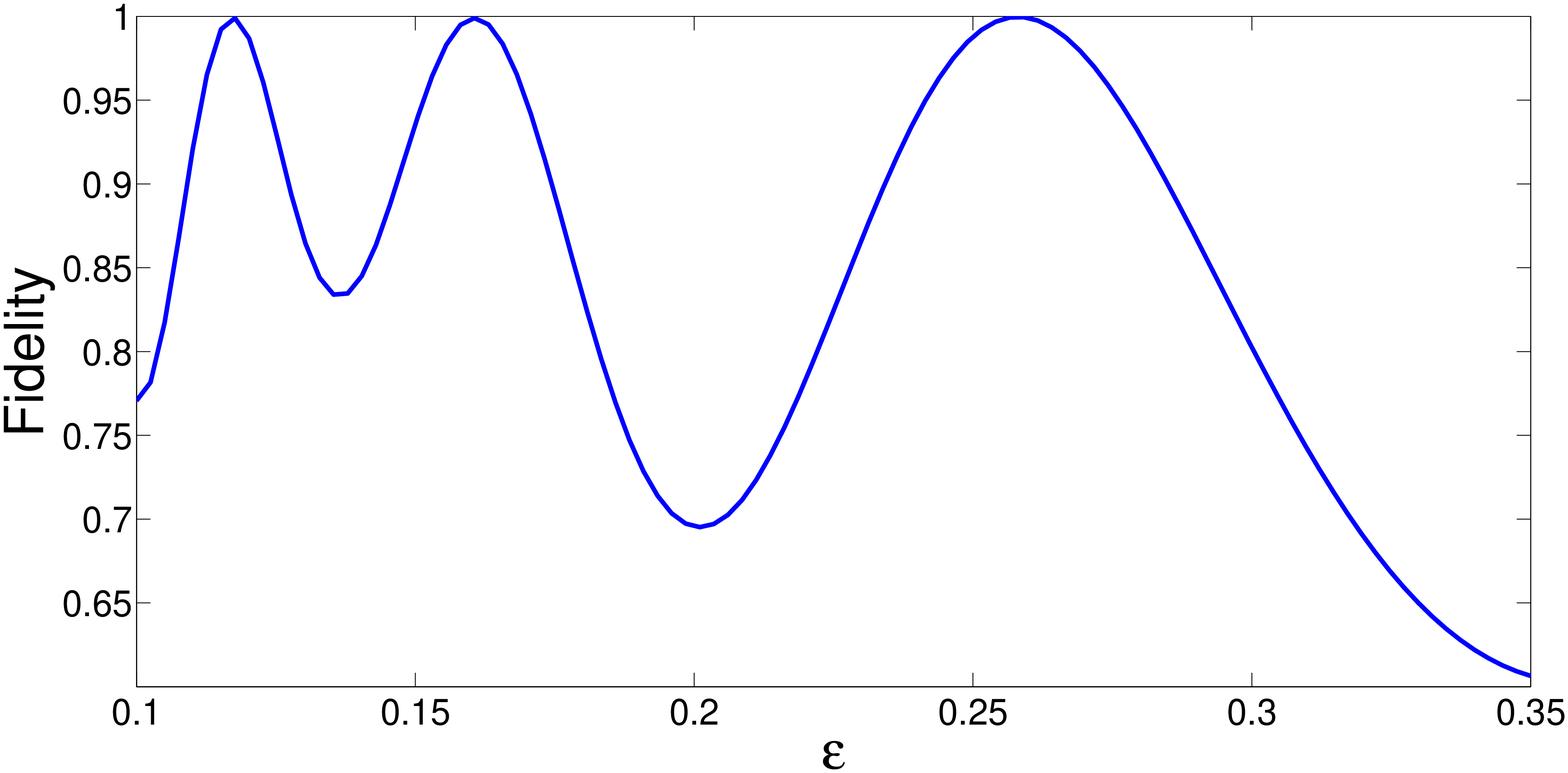}}
 \caption{The fidelity $F_{1}$ versus the parameter $\epsilon$.}
 \label{FE}
\end{figure}

\begin{figure}
 \renewcommand\figurename{\small FIG.}
 \centering \vspace*{8pt} \setlength{\baselineskip}{10pt}
 \subfigure[]{
 \includegraphics[scale = 0.17]{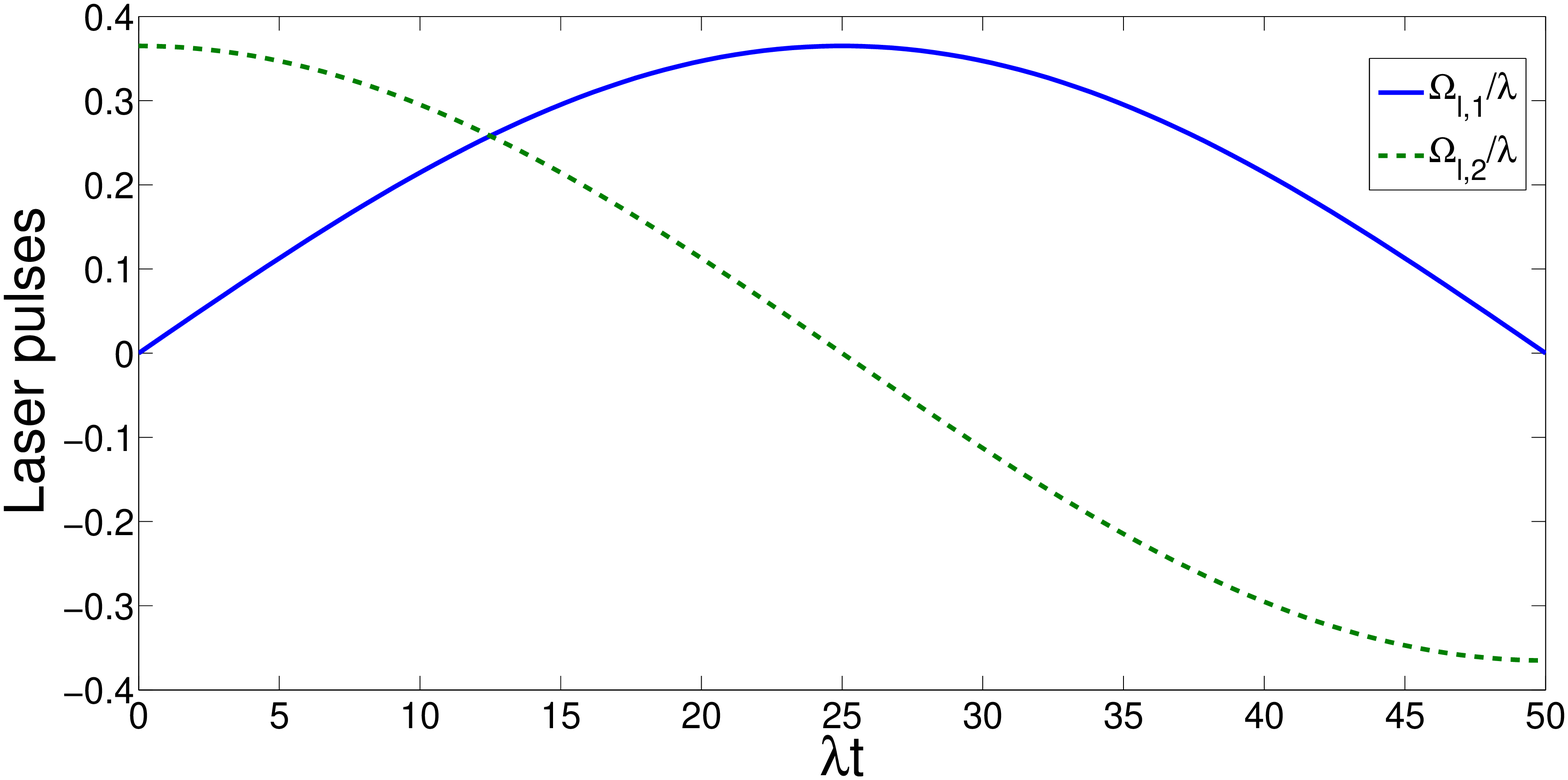}}
 \subfigure[]{
 \includegraphics[scale = 0.17]{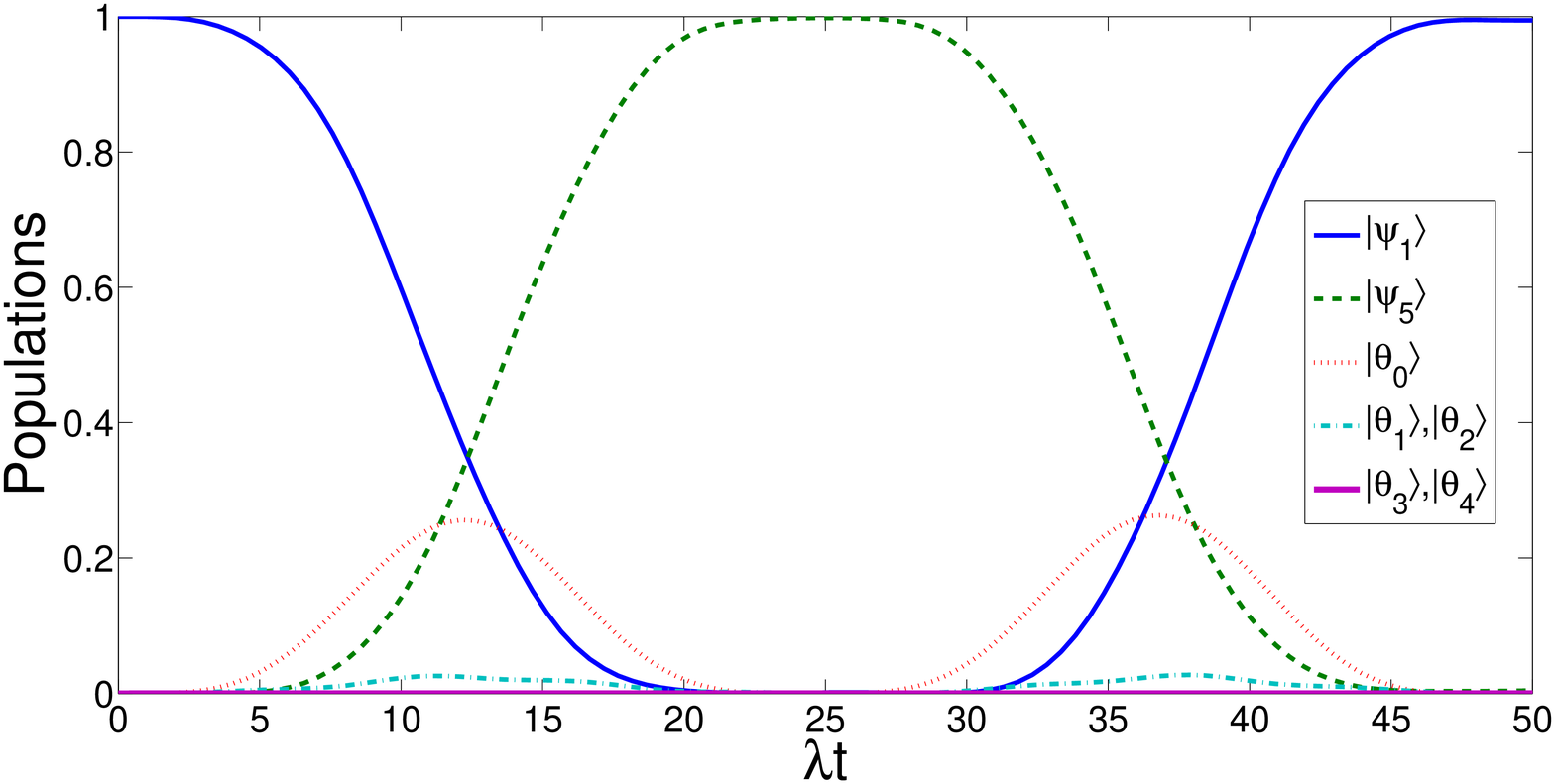}}
 \caption{
    (a) The time dependence of the laser fields $\Omega_{l,1}$ and $\Omega_{l,2}$ when $v=2$ and $\epsilon=0.258$.
    (b) Time evolution of the populations for the states $|\psi_{1}\rangle$, $|\psi_{5}\rangle$, and $|\theta_{m}\rangle$ ($m=0,1,\cdots,5$).
          }
 \label{O1O2P01234}
\end{figure}

\begin{figure}
 \renewcommand\figurename{\small FIG.}
 \centering \vspace*{8pt} \setlength{\baselineskip}{10pt}
 \subfigure[]{
 \includegraphics[scale = 0.17]{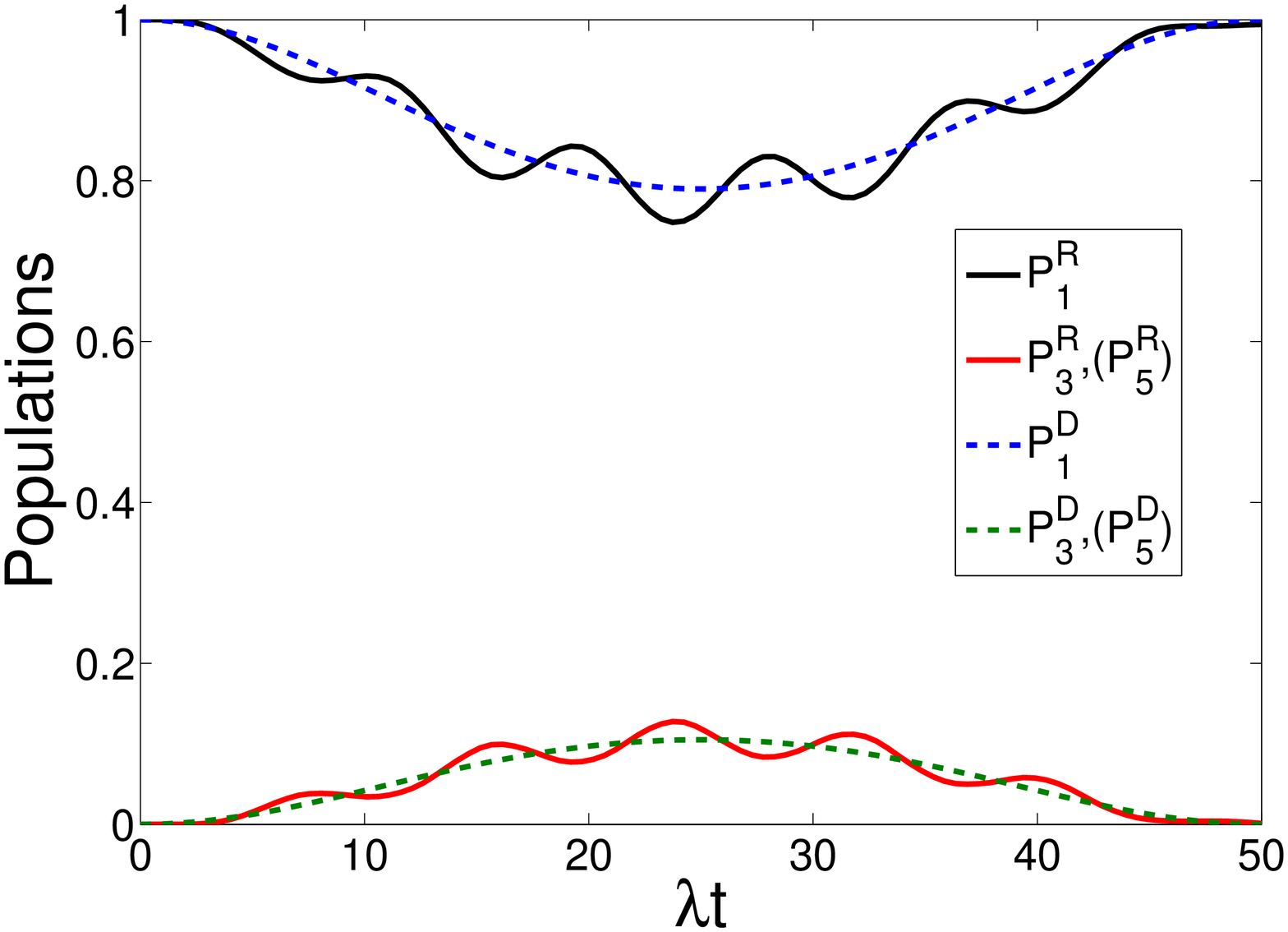}}
 \subfigure[]{
 \includegraphics[scale = 0.17]{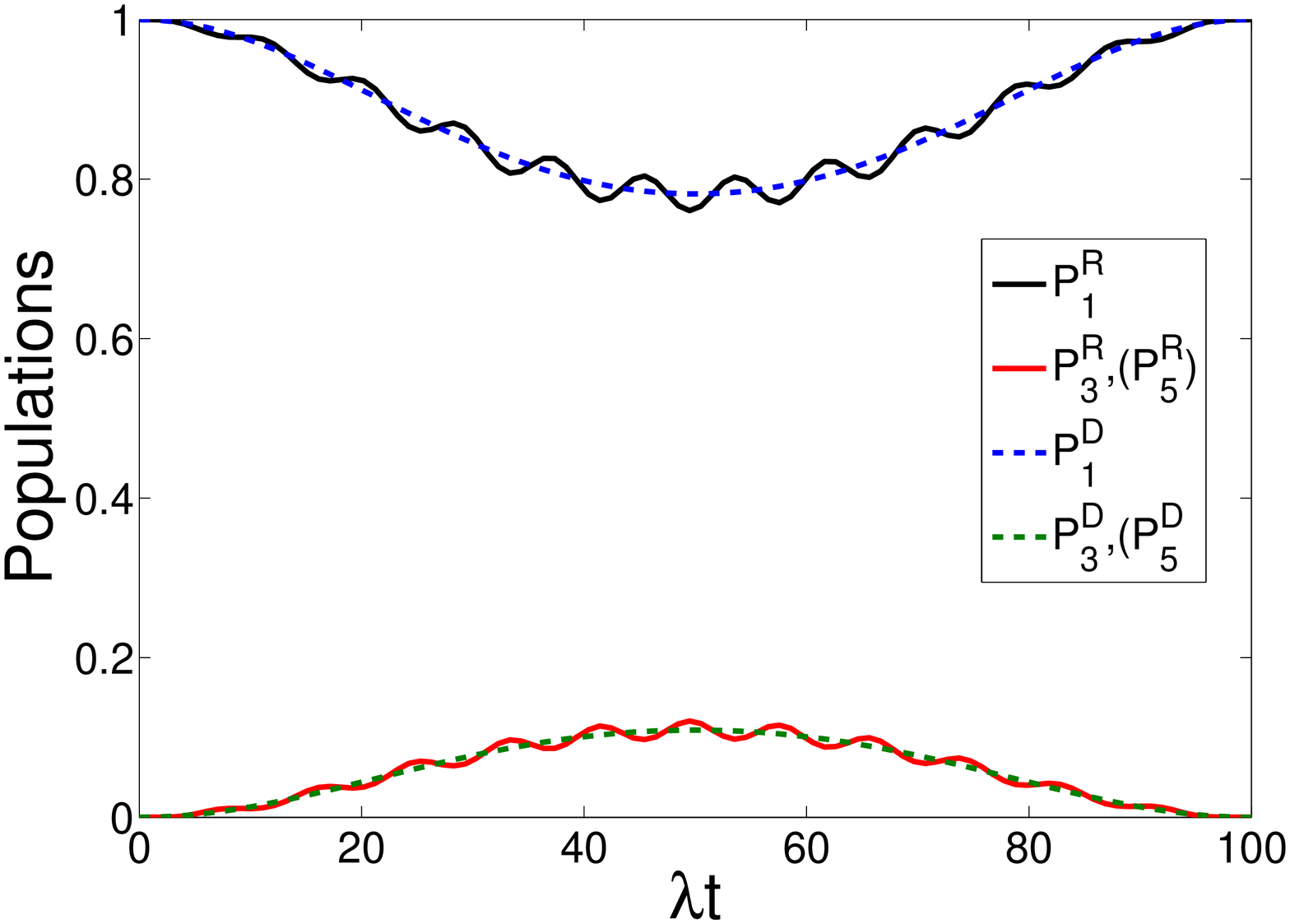}}
 \subfigure[]{
 \includegraphics[scale = 0.17]{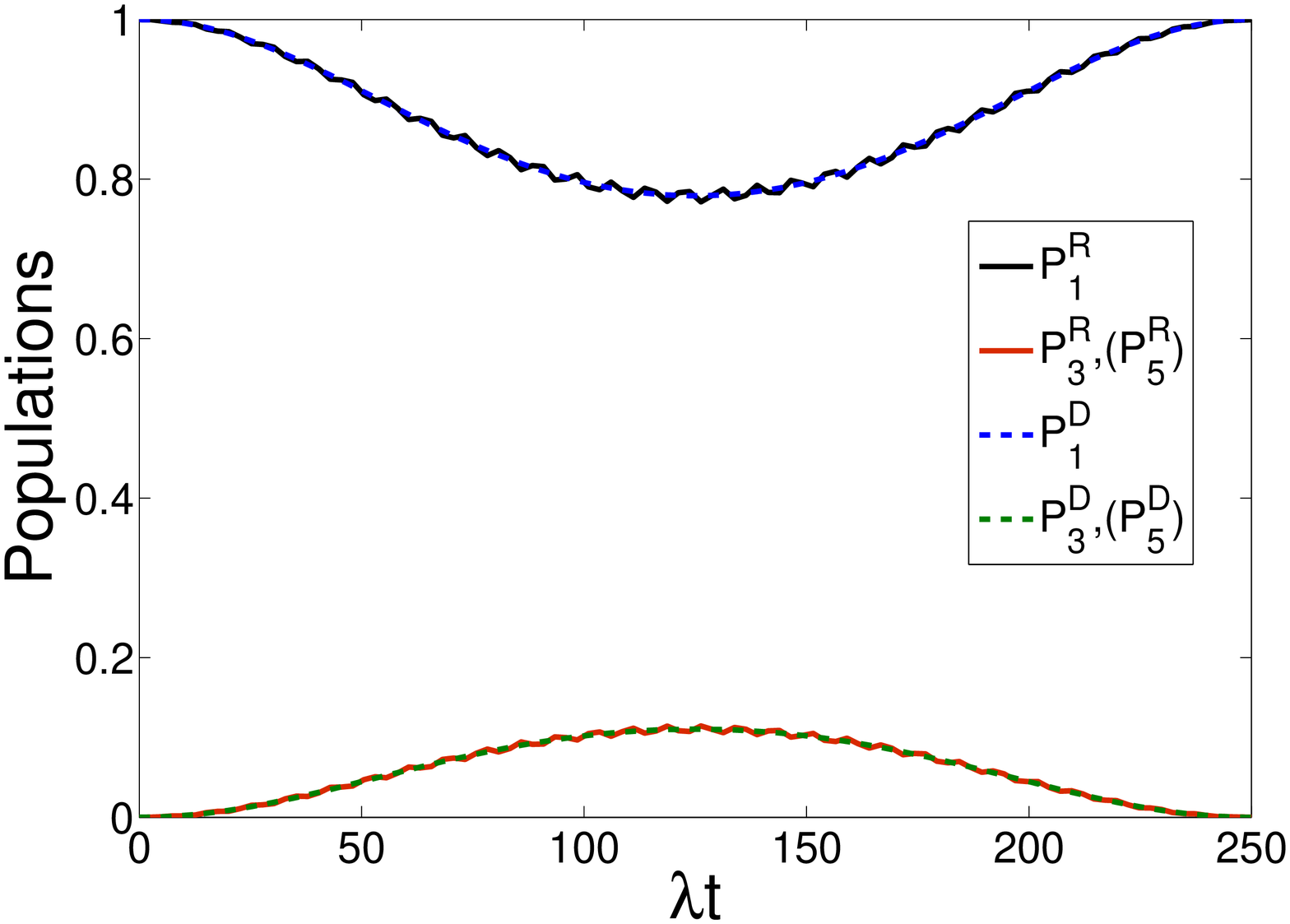}}
 \caption{
     Time evolution of the populations for the states $|\varphi_{1}\rangle$, $|\varphi_{3}\rangle$, and $|\varphi_{5}\rangle$ in different cases:
     (a) $\zeta=1$ and $t_{f}=50/\lambda$, (b) $\zeta=2$ and $t_{f}=100/\lambda$, (c) $\zeta=5$ and $t_{f}=250/\lambda$.
        }
 \label{FsD}
\end{figure}

The validity of the above theoretical analysis will be numerically
proved in the following. From eq. (\ref{eq3-10}), we know that the amplitude
of the laser pulses is
\begin{eqnarray}\label{eq5-1}
  \Omega_{0}=\frac{\chi \pi \cot{\epsilon}}{vt_{f}}=\frac{\pi \cot{\epsilon}}{t_{f}}\sqrt{2+(\frac{\lambda}{v})^{2}}.
\end{eqnarray}
That is, the amplitude of the laser pulses depends on three parameters: $\epsilon$, $t_{f}$, and $v$.
And there is an inverse relationship between $\Omega_{0}$ and $t_{f}$ ($v$) when $\epsilon$ is a constant value. Fig. \ref{Fvtf} (a) shows the fidelity of
the target state $-|\psi_{1}\rangle$ versus parameters $\lambda t_{f}$ and $v/\lambda$ in case of $\epsilon=\arcsin{0.25}$. From the
figure, we find that, when $t_{f}=20/\lambda$ and $v=0.7\lambda$, the fidelity is also higher than $90\%$. While with these two
parameters, the laser amplitude is $\Omega_{0}=1.22$ which does not meet the Zeno condition $\Omega_{l,k}\ll \lambda$.
The reason for this result has been discussed in refs. \cite{YHCYXQQCJSPRA14,YHCYXQQCJSLPL14} in very detail: the population transfer form the initial state $|\psi_{1}\rangle$ to
the target state $|\psi_{7}\rangle$ can be fast and perfectly achieved even when the Zeno condition has not been fulfilled and vice versa. However,
in this paper, we focus on how to use the LR phases to implement a $\pi$ phase gate. From the theoretical analysis above,
we know that when the Zeno condition is destroyed, the phase change of the whole system is unpredictable and unanalyzable. Therefore, through
the relationship in eq. (\ref{eq5-1}), we plot Fig. \ref{Fvtf} (b) in case of $\epsilon=\arcsin{0.25}$ to accurately choose $v$ and $t_{f}$ to make sure the Zeno
condition is fulfilled. The result shows that when $t_{f}$ or $v$ is large enough, $\Omega_{0}$ only changes a little with the changing of $t_{f}$ or $v$.
To complete the gate operation in a short interaction time, it seems that $t_{f}=50/\lambda$ and $v=2\lambda$ is a good choice for the scheme.
The above analysis is based on $\epsilon$ is a constant value, while,
we know there are also some different choices for the parameter $\epsilon$ as shown in Fig. \ref{FE}.
Figure \ref{FE} is plotted with $\{t_{f}=50/\lambda,\ v=2\lambda\}$, and we find that the ideal value of $\epsilon$
for the highest fidelity is slightly different from the condition $\epsilon=\arcsin{1/\zeta}$, the reason for this
difference has been discussed in ref. \cite{YHCYXQQCJSPRA14} in detail: in the present case, the Zeno condition is satisfied but not very ideally because
speeding up the system requires relatively large laser intensities.
Therefore, under the premise that the interaction time for the gate operation is short,
to satisfy the Zeno condition as well as possible, the parameters for $\Omega_{0}$ is chosen as
$\{\epsilon=0.258,\ v=2\lambda,\ t_{f}=50/\lambda\}$. With these parameters, as shown in Fig. \ref{O1O2P01234} (a),  the laser pulses are determined.
And the time evolution of the populations for states $|\psi_{1}\rangle$, $|\psi_{5}\rangle$, $|\theta_{0}\rangle$,
$|\theta_{1}\rangle$ ($|\theta_{2}\rangle$), and $|\theta_{3}\rangle$ ($|\theta_{4}\rangle$)
are shown in Fig. \ref{O1O2P01234} (b). An obvious phenomenon in the figure which is much different from
the adiabatic passage technique is that the intermediate state $|\theta_{0}\rangle$ is populated during the evolution
(in a scheme based on adiabatic passage, the state $|\theta_{0}\rangle$ is usually neglected).
We can understand this from refs. \cite{XCILARDGJGMPrl10,XCJGMPra12,YHCYXQQCJSPRA14},
the key point to speed up the population transfer is increasing the intermediate states, especially, the states which are
neglected during the adiabatic passage. Ideally, the more the intermediate states are populated, the faster the evolution is. However,
a high population of an intermediate state usually leads to decoherence of the system. What is more drawn from Fig. \ref{O1O2P01234} (b)
is that we can confirm the Zeno condition is fulfilled with this
set of parameters: the states $|\theta_{1}\rangle$, $|\theta_{2}\rangle$, $|\theta_{3}\rangle$ and $|\theta_{4}\rangle$
are all considered as negligible during the evolution. In Fig. \ref{FsD} (a), we show the evolution of the system when the initial state
is $|\varphi_{1}\rangle$. The solid curves which are plotted through the relation $P^{R}_{j}=|\langle \varphi_{j}|\rho_{s}(t)|\varphi_{j}\rangle|$
($j=1,3,5$) are the time-dependent populations for states $|\varphi_{j}\rangle$, and the dashed curves which are plotted through the relation
$P^{D}_{j}=|\langle\varphi_{j}|\tilde{\theta}_{0}(t)\rangle|^{2}$ are also the time-dependent populations for states $|\varphi_{j}\rangle$,
where $\rho_{s}(t)$ which satisfies $\partial_{t}{\rho_{s}}=i[\rho_{s},H_{i,s}]$
is the density operator for the Hamiltonian $H_{i,s}$ in eq. (\ref{eq3-13a}) and $|\tilde{\theta}_{0}(t)\rangle$ is the dark state in eq. (\ref{eq3-13}).
From this figure, we find there are some differences between the evolutions governed by $H_{i,s}$ and the dark state: it seems that
the dark state can not ideally describe the evolution of the system. Through analyzing
refs. \cite{XCJGMPra12,YHCYXQQCJSPRA14,YHCYXQQCJSLPL14,TALSSPra96} in detail, a favorable inference is drawn:
for a resonant $\Lambda$-type three-level system as shown in ref. \cite{TALSSPra96}, if the stokes and pump pulses are designed with the form
\begin{eqnarray}\label{eq5-2}
  \Omega_{p}(t)=\Omega_{0}\sin{\frac{\beta}{2}}, \
  \Omega_{s}(t)=\Omega_{0}\cos{\frac{\beta}{2}},
\end{eqnarray}
$\zeta=(\sin{\epsilon})^{-1}/{4}$ seems to dominate
the oscillation number of the curves which describe the time-dependent populations of the three states.
When $\zeta$ is too small, for instance, $\zeta=1,2$, the system can not meet the adiabatic condition.
Numerical simulation shows that, in this $\Lambda$-type system the oscillation number $N_{o}$ equals to $\zeta$.
And we draw an inference that at least under certain conditions, for an adiabatic system, the oscillation number is
in direct proportion to $\zeta$. For fair comparison and to make sure the adiabatic condition for the Hamiltonian in eq. (\ref{eq3-13a}) is fulfilled,
it is better to make the population for the same state, i.e., the state $|\varphi_{1}\rangle$ keeps unchanging with different parameters.
We choose the population of state $|\varphi_{1}\rangle$ as a typical example for the analysis. From eq. (\ref{eq3-13}), the population
of state $|\varphi\rangle$ is given by
\begin{eqnarray}\label{eq5-3}
  P_{1}^{D}=\frac{\lambda^{2}}{2\Omega_{l,1}^{2}+\lambda^{2}}=\frac{\lambda^{2}}{2\Xi^{2}(\frac{\cot{\epsilon}}{t_{f}})^{2}+\lambda^{2}},
\end{eqnarray}
where $\Xi=(\pi\chi\sin{\beta})/v$. To keep $P_{1}^{D}$ the same when we change $\zeta$, we change $t_{f}$ accordingly.
$(\cot{\epsilon})/t_{f}=C$ has to be satisfied, where $C$ is a constant. And we obtain
\begin{eqnarray}\label{eq5-4}
  \frac{\cot{\epsilon}}{t_{f}}=C\Rightarrow t_{f}=\frac{\cot{\epsilon}}{C}=\frac{\sqrt{1-(\frac{1}{4\zeta})^{2}}}{4C\zeta}\simeq\frac{1}{4C\zeta}.
\end{eqnarray}
The result shows there is an inverse relationship between the total interaction time $t_{f}$ and $\zeta$. Therefore, when we
choose $\zeta=2$, for fair comparison, it is better to choose $t_{f}=100/\lambda$.
In Fig. \ref{FsD} (b), we plot $P^{R}_{j}$ and $P^{D}_{j}$ again with parameters $\{\zeta=2,t_{f}=100/\lambda,v=2\lambda\}$.
Contrasting Fig. \ref{FsD} (a) with (b), we find that when $\zeta=2$, the oscillation number is double of that when $\zeta=1$,
and the more the oscillation number is, the smoother the curve is. When we choose $\zeta=5$ and $t_{f}=250/\lambda$, the
curves $P_{j}^{R}$ coincide exactly with curves $P_{j}^{D}$ as shown in Fig. \ref{FsD} (c),
that is to say, the dark state can describe the evolution governed by $H_{i,s}$ ideally
with a relative large $\zeta$. However, a large $\zeta$ leads to a small laser amplitude $\Omega_{0}$, and a small laser amplitude will
lead to a long interaction time which is a deviation from our intention.

\begin{figure}
 \renewcommand\figurename{\small FIG.}
 \centering \vspace*{8pt} \setlength{\baselineskip}{10pt}
 \subfigure[]{
 \includegraphics[scale = 0.17]{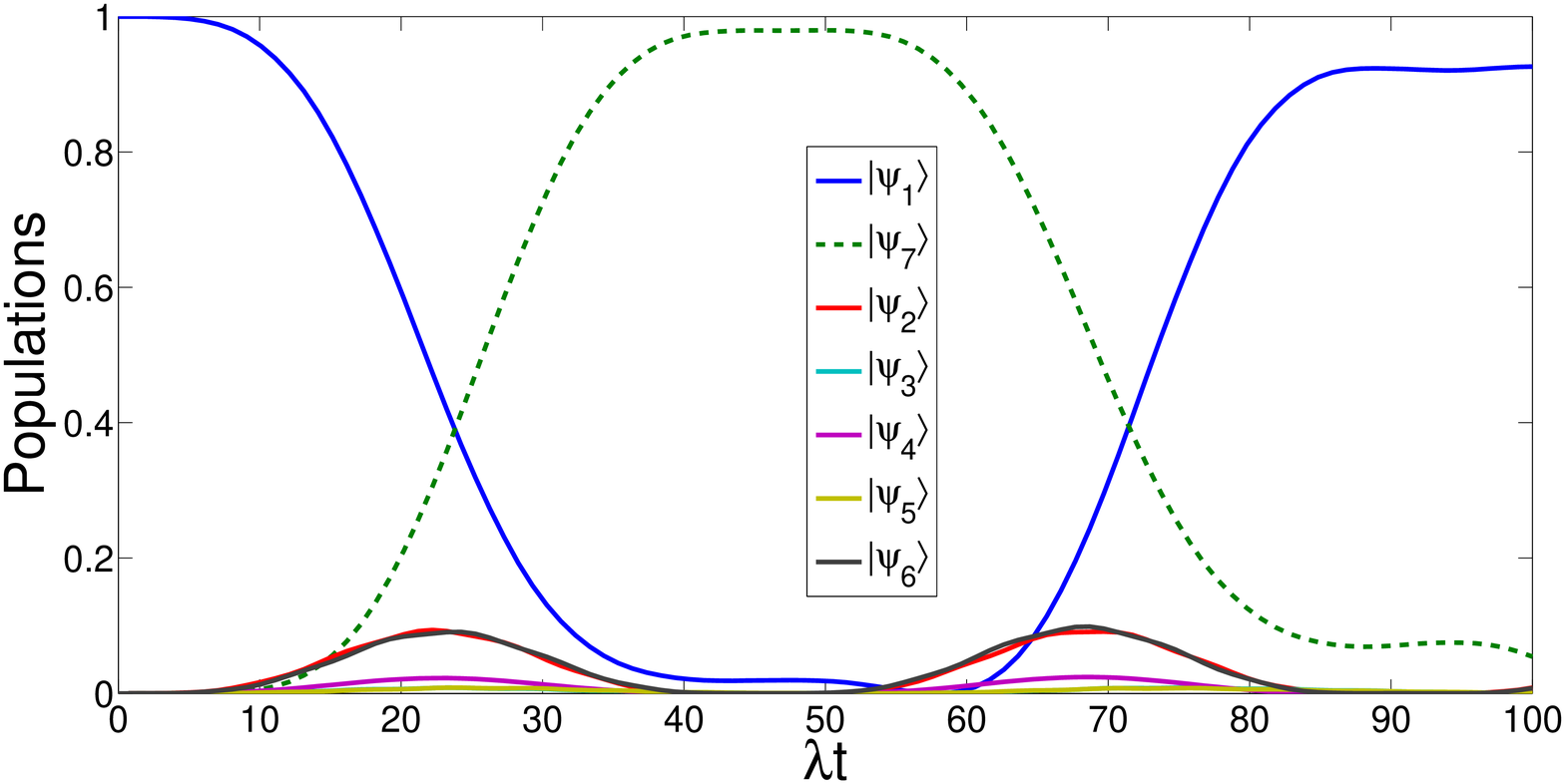}}
 \subfigure[]{
 \includegraphics[scale = 0.17]{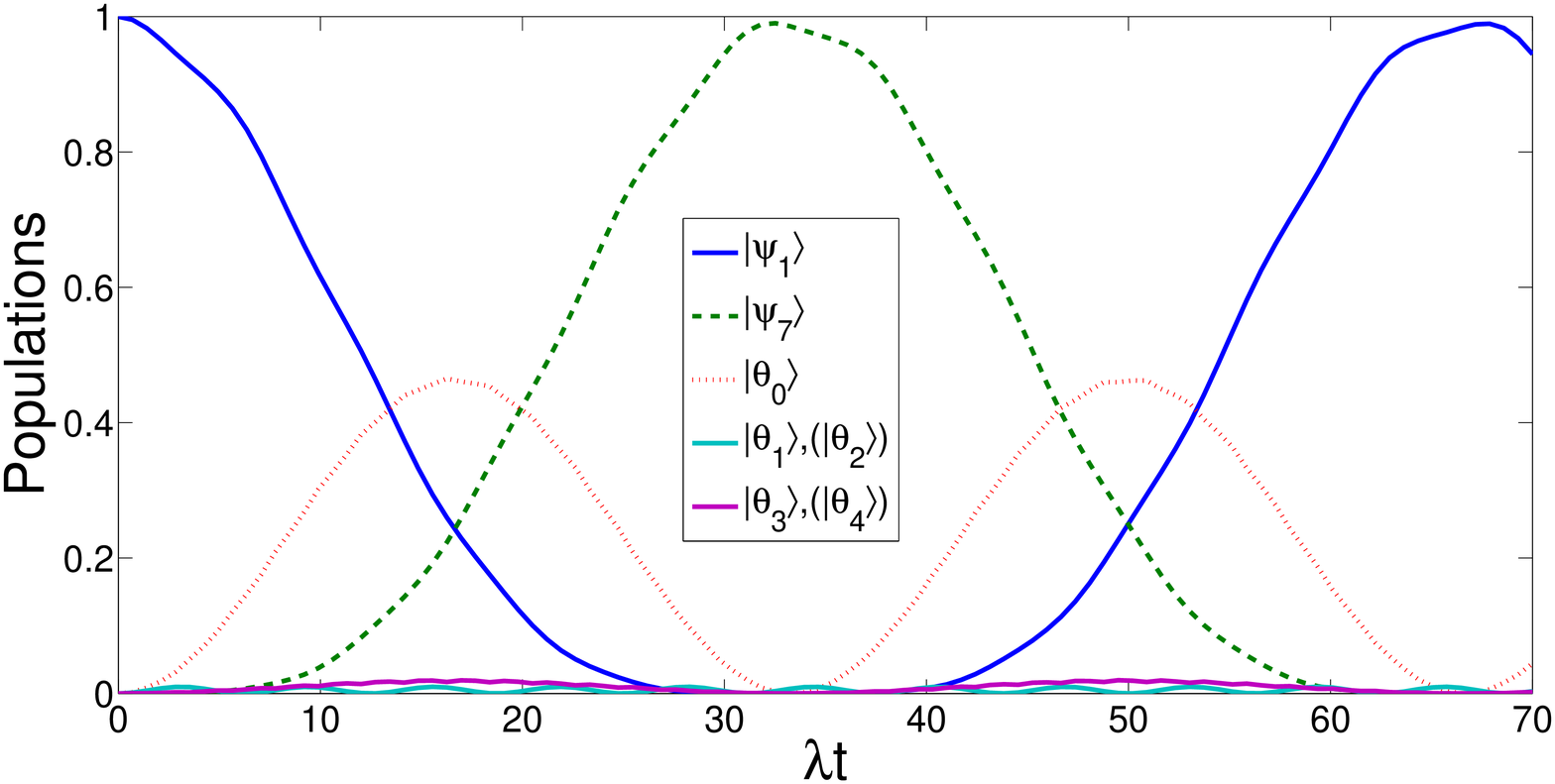}}
 \caption{
     Time-dependent evolution of similar schemes for the $\pi$ phase gates:
    (a) via adiabatic passage.
    (b) via quantum Zeno dynamics.
          }
 \label{FadiZeno}
\end{figure}

We contrast the present scheme with adiabatic and Zeno schemes respectively to prove that STAP has effectively shortened the interaction time. A $\pi$ phase gate is achieved via adiabatic evolution of dark eigenstates or quantum Zeno dynamics in this two-atom system. The Hamiltonian takes the form in eq. (\ref{eq3-1}), we only have to change the parameters for the laser pulses.
By solving the characteristic equation of $H_{i}$ the dark state is given by
\begin{eqnarray}\label{eq5-5}
  |Dark(t)\rangle=\frac{1}{N_{D}}[\Omega_{l,2}|\psi_{1}\rangle-\Omega_{l,1}|\psi_{7}\rangle
                              -\frac{\Omega_{l,1}\Omega_{l,2}}{\lambda}(|\psi_{3}\rangle-|\psi_{5}\rangle)].
\end{eqnarray}
For simplicity, we also choose the form of the laser pulses as
$\Omega_{l,1}=\Omega_{0}'\sin{\beta}$ and $\Omega_{l,2}=\Omega_{0}'\cos{\beta}$ to meet the boundary conditions for stimulated Raman scattering involving adiabatic passage.
When $t=t_{f}$, $\Omega_{l,1}=-\Omega_{0}'$ and $\Omega_{l,2}=0$, and
the final state becomes $|Dark(t_{f})\rangle=-|\psi_{1}\rangle$. Since the adiabatic condition $|\langle Dark|\partial_{t}\Phi_{m\neq0}\rangle|\ll |E_{m}|$
is too complex to directly analyze, where
$|\Phi_{m}\rangle$ is the $m$th instantaneous eigenstate corresponding nonzero eigenvalue $E_{m}$, we simplify the whole system
by setting a limit condition $\Omega_{l,1},\Omega_{l,2}\ll \lambda,v$ similar to what we do in section III. Then the Hamiltonian can be
simplified as the form in eq. (\ref{eq3a-5}), and the adiabatic condition is also simplified as
$|\langle \eta_{0}|\partial_{t}\eta_{\pm}\rangle|\ll|\varepsilon_{\pm}|$. For the given laser pulses, we have
\begin{eqnarray}\label{eq5-6}
  \frac{\dot{\beta}}{\sqrt{2}}\ll\frac{v\Omega_{0}'}{\chi}\Rightarrow t_{f}\gg \frac{\pi\chi}{\sqrt{2}v\Omega_{0}'}.
\end{eqnarray}
When we choose $v=2\lambda$, $(\pi\chi)/({\sqrt{2}v\Omega_{0}'})\approx(\Omega_{0}')^{-1}$. Therefore, the interaction time $t_{f}$ must be chosen at least $20$ times larger than $(\Omega_{0}')^{-1}$ to obtain a high fidelity of the
target state.
To prove this, we plot Fig. \ref{FadiZeno} (a) which shows the time evolution of the populations for states $|\psi_{1}\rangle-|\psi_{7}\rangle$
when $\Omega_{0}=0.2\lambda$ and $t_{f}=100/\lambda=20(\Omega_{0}')^{-1}$.
Shown in the figure, even with $t_{f}=20(\Omega_{0}')^{-1}$, the adiabatic condition is not ideally fulfilled, the fidelity of the target state
$-|\psi_{1}\rangle$ is only about $92\%$ and states $|\psi_{2}\rangle$
and $|\psi_{6}\rangle$ which should have been adiabatically eliminated are populated. That means,
if we want to achieve an ideal target state $-|\psi_{1}\rangle$,
the interaction time required in the adiabatic scheme is longer than $20(\Omega_{0}')^{-1}$. And in a scheme also with Hamiltonian $H_{i}$
to implement a $\pi$ phase gate via quantum Zeno dynamics, when the Zeno condition is fulfilled, the evolution of the system is
approximatively governed by the effective Hamiltonian in eq. (\ref{eq3a-5}), and the general evolution form of eq. (\ref{eq3a-5}) in time $t$ is
\begin{eqnarray}\label{eq5-7}
  |\psi(t)\rangle&=&\frac{v^2}{(\chi\theta)^{2}}(\Omega_{l,1}^{2}\cos{\theta t}+\Omega_{l,2}^{2})|\psi_{1}\rangle
                   -i\sin{\theta t}|\theta_{0}\rangle                                                                 \cr\cr
                  &&+\frac{v^{2}}{(\chi\theta)^{2}}(\Omega_{l,1}\Omega_{l,2}\cos{\theta t}-\Omega_{l,1}\Omega_{l,2})|\psi_{7}\rangle,
\end{eqnarray}
where $\theta=\sqrt{v^{2}(\Omega_{l,1}^{2}+\Omega_{l,2}^{2})/\chi^{2}}$. When we choose $t_{f}=2\pi/\theta$
and $\Omega_{1}=\Omega_{2}=i\Omega_{0}'$ ($\Omega_{0}'$ is a real number), the
final state becomes $|\psi(t_{f})\rangle=-|\psi_{1}\rangle$. As known to all, the smaller the value of $\Omega_{0}'$ is,
the better the Zeno condition is satisfied. Whereas, a smaller $\Omega_{0}'$ leads to a longer operation time.
Therefore, we choose a relative large $\Omega_{0}'=0.1\lambda$ for the Zeno scheme. In this case, the operation time is
$t_{f}=2\pi/\theta\approx21.2\pi/\lambda$ which is a little longer than that of the shortcut scheme
when $v=2\lambda$ as shown in Fig. \ref{FadiZeno} (b). With these parameters, the fidelity of
the target state is about $98.9\%$. It is notable that the state $|\theta_{0}\rangle$ including atomic excited states is greatly populated
which makes the scheme much more sensitive to decoherence caused by atomic spontaneous emission than the shortcut scheme. In general, to avoid this defect,
researchers usually create a detuning between the effective transition $|\psi_{1}\rangle(|\psi_{7}\rangle)\leftrightarrow|\theta_{0}\rangle$ in theory.
However, this operation inevitably increases the operation time of the scheme. Limited by the Zeno condition, the present shortcut scheme might
not have distinct advantage in operation time compared with the Zeno scheme, but when all of the merits and faults are considered together,
the shortcut scheme is much better than the other two without doubt.

\begin{figure}
 \renewcommand\figurename{\small FIG.}
 \centering \vspace*{8pt} \setlength{\baselineskip}{10pt}
 \subfigure[]{
 \includegraphics[scale = 0.17]{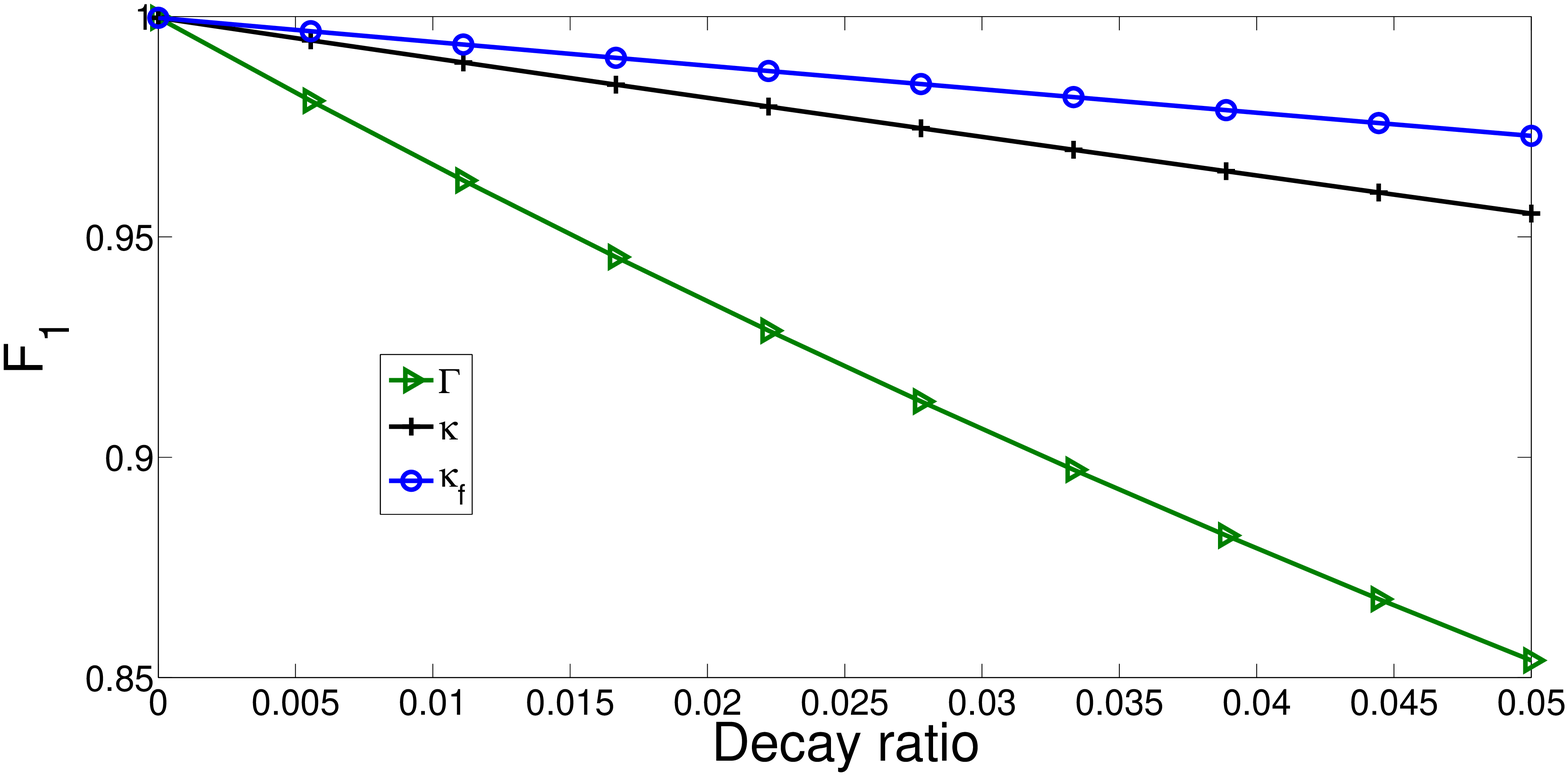}}
 \subfigure[]{
 \includegraphics[scale = 0.17]{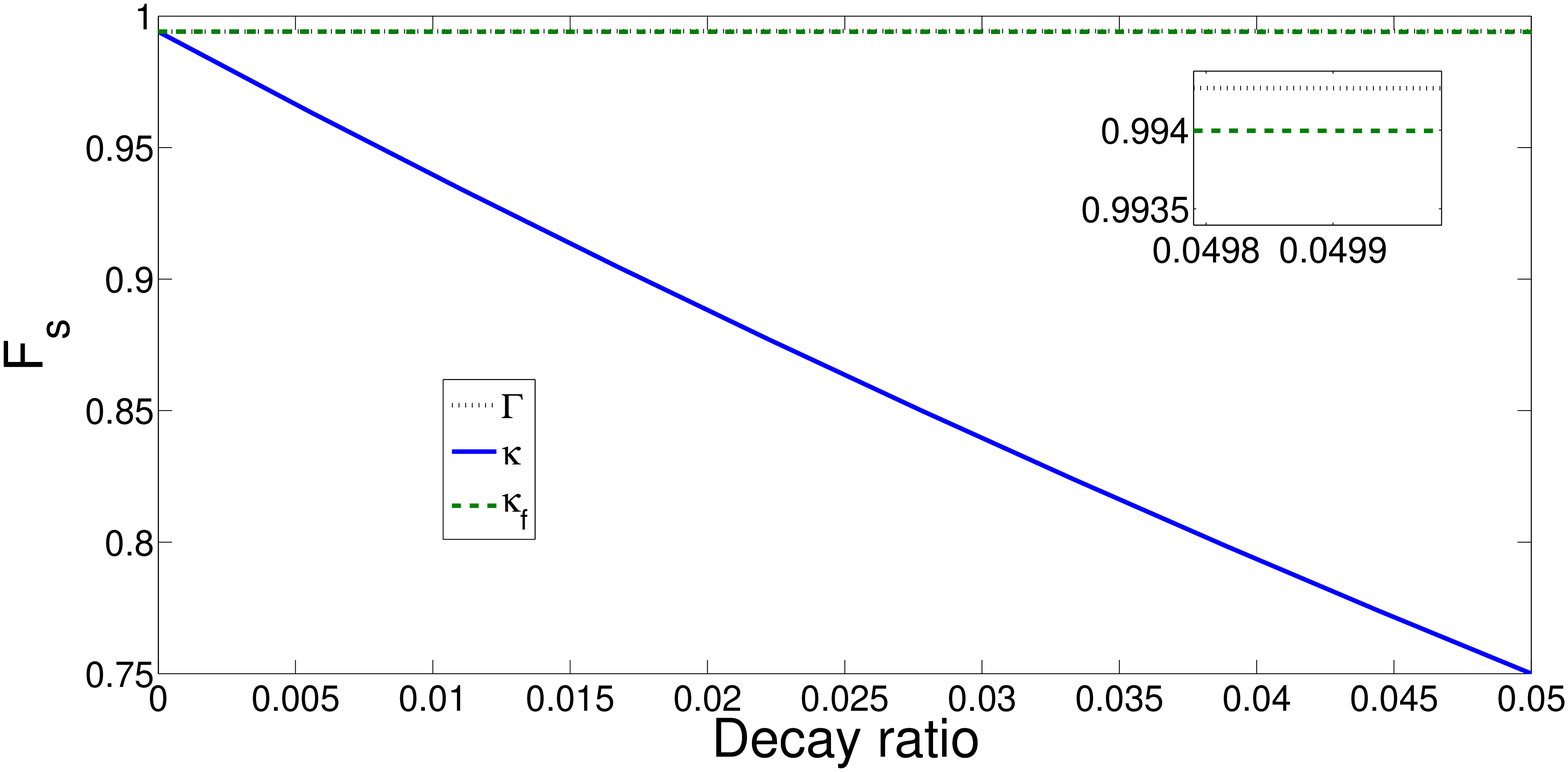}}
 \caption{
    (a) The fidelity $F_{1}$ of the target state $-|\psi_{1}\rangle$ versus the three noise resources $\Gamma$, $\kappa$, and $\kappa_{f}$.
    (b) The fidelity $F_{s}$ of the target state $|\psi_{s}\rangle$ versus the three noise resources $\Gamma$, $\kappa$, and $\kappa_{f}$.
          }
 \label{Fkrkf}
\end{figure}

Robustness against possible mechanisms of decoherence is important for a scheme to be
applicable in quantum-information processing and quantum computing. It is necessary for us
to examine robustness of our shortcut schemes described in the previous sections against decoherence
mechanisms including the atomic spontaneous emission, the cavity decay, and the fiber decay. Once the decoherence is considered,
the evolution of the system can be modeled by a master equation in Lindblad form,
\begin{eqnarray}\label{eq5-8}
 \dot{\rho}=i[\rho,H_{i}]+\sum_{k}{[L_{k}\rho L_{k}^{\dag}-\frac{1}{2}(L_{k}^{\dag}L_{k}\rho+\rho L_{k}^{\dag}L_{k})]},
\end{eqnarray}
where $\rho$ is the density operator for the whole system, $L_{k}$ are the so-called Lindblad operators. Firstly,
we consider the initial state is $|\psi_{1}\rangle$. For the implementation of the $\pi$ phase gate,
there are seven Lindblad operators governing dissipation in the model when the initial state is $|\psi_{1}\rangle$:
\begin{eqnarray}\label{eq5-9}
  L_{1}^{\Gamma}&=&\sqrt{\Gamma_{1}}|f\rangle_{1}\langle e|,\
  L_{2}^{\Gamma}=\sqrt{\Gamma_{2}}|f\rangle_{2}\langle e|,\
  L_{3}^{\Gamma}=\sqrt{\Gamma_{3}}|g\rangle_{1}\langle e|,\
  L_{4}^{\Gamma}=\sqrt{\Gamma_{4}}|g\rangle_{2}\langle e|,\ \cr
  L_{5}^{\kappa_{c}}&=&\sqrt{\kappa_{1}}a_{1},\
  L_{6}^{\kappa_{c}}=\sqrt{\kappa_{2}}a_{2},\
  L_{7}^{\kappa_{f}}=\sqrt{\kappa_{f}}b.
\end{eqnarray}
where $\Gamma_{i}$ ($i=1,2,3,4$) are the atomic spontaneous emissions, $\kappa_{j}$ ($j=1,2$) are the cavity decays, and $\kappa_{f}$ is the fiber decay.
For simplicity, we set $\Gamma_{j}=\Gamma/2$ and $\kappa_{i}=\kappa$. In Fig. \ref{Fkrkf} (a), we plot the fidelity $F_{1}$ of the
target state $-|\psi_{1}\rangle$ through the relation $F_{1}=|\langle -\psi_{1}|\rho(t_{f})|-\psi_{1}\rangle|$
versus the dimensionless parameters $\Gamma/\lambda$, $\kappa/\lambda$ and $\kappa_{f}/\lambda$
via numerically solving the master equation (\ref{eq5-8}).
We can find from the figure that the method is robust against the cavity decay and fiber decay while sensitive to the atomic spontaneous
emission. This result can be understood with the help of Fig. \ref{O1O2P01234} (b): the state $|\theta_{0}\rangle$
including atomic excited states $|\psi_{2}\rangle$ and $|\psi_{4}\rangle$ is populated, while the states $|\theta_{m}\rangle$ ($m=1,2,3,4$)
including cavity- and fiber-excited states keep ignorable during the evolution.
Then, if the initial state is $|\varphi_{1}\rangle$, there will be five Lindblad operators governing dissipation:
\begin{eqnarray}\label{eq5-10}
  L_{1}^{\Gamma}=\sqrt{\Gamma_{1}}|f\rangle_{1}\langle e|,\
  L_{2}^{\Gamma}=\sqrt{\Gamma_{2}}|f\rangle_{2}\langle e|,\
  L_{3}^{\kappa_{c}}=\sqrt{\kappa_{1}}a_{1},\
  L_{4}^{\kappa_{c}}=\sqrt{\kappa_{2}}a_{2},\
  L_{5}^{\kappa_{f}}=\sqrt{\kappa_{f}}b.
\end{eqnarray}
We display the dependence on the ratios $\Gamma/\lambda$, $\kappa/\lambda$, and $\kappa_{f}/\lambda$
of the fidelity of the final state $|\varphi_{1}\rangle$ through solving the corresponding master equation in Fig. \ref{Fkrkf} (b).
The result is much different from that of $F_{1}$, when the initial state is $|\varphi_{1}\rangle$, the atomic spontaneous emissions
and the fiber decay almost have no influence on the fidelity while the decay of cavity has great influence.
It is because when the initial state is $|\varphi_{1}\rangle$, the adiabatic condition is fulfilled and the
states $|\varphi_{2}\rangle$ including atomic excited state and $|\varphi_{4}\rangle$ including fiber-excited
state have been effectively adiabatically eliminated. Whereas, shorting the operation time requires
a relatively large pulse amplitude, so the intermediate states $|\varphi_{3}\rangle$
and $|\varphi_{5}\rangle$ in the dark state $|\tilde{\theta}_{0}\rangle$ including the excited states of the cavity can not
be effectively eliminated during the evolution, and that makes the system sensitive to the cavity decay.

\begin{figure}
 \renewcommand\figurename{\small FIG.}
 \centering \vspace*{8pt} \setlength{\baselineskip}{10pt}
 \subfigure[]{
 \includegraphics[scale = 0.23]{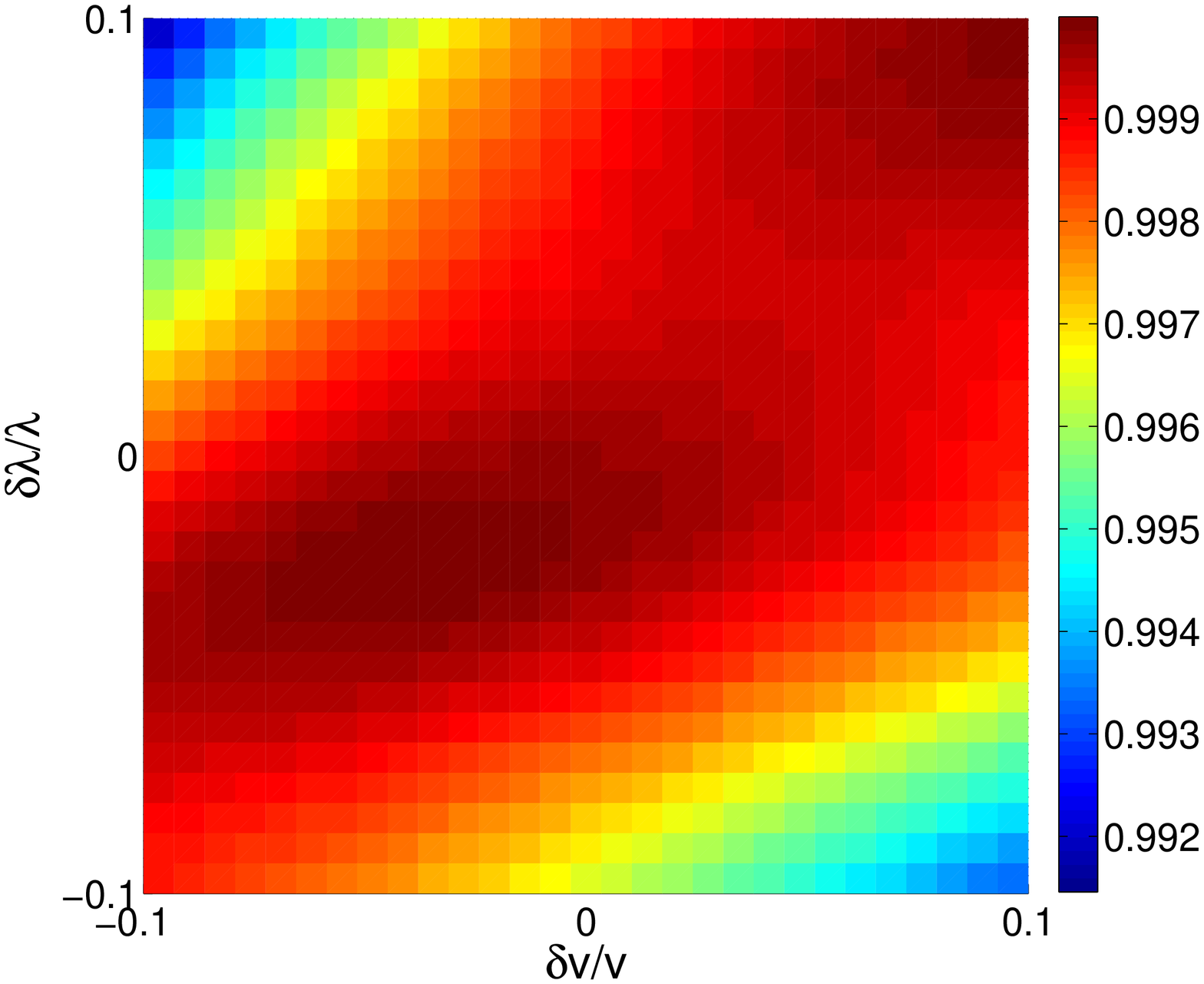}}
 \subfigure[]{
 \includegraphics[scale = 0.23]{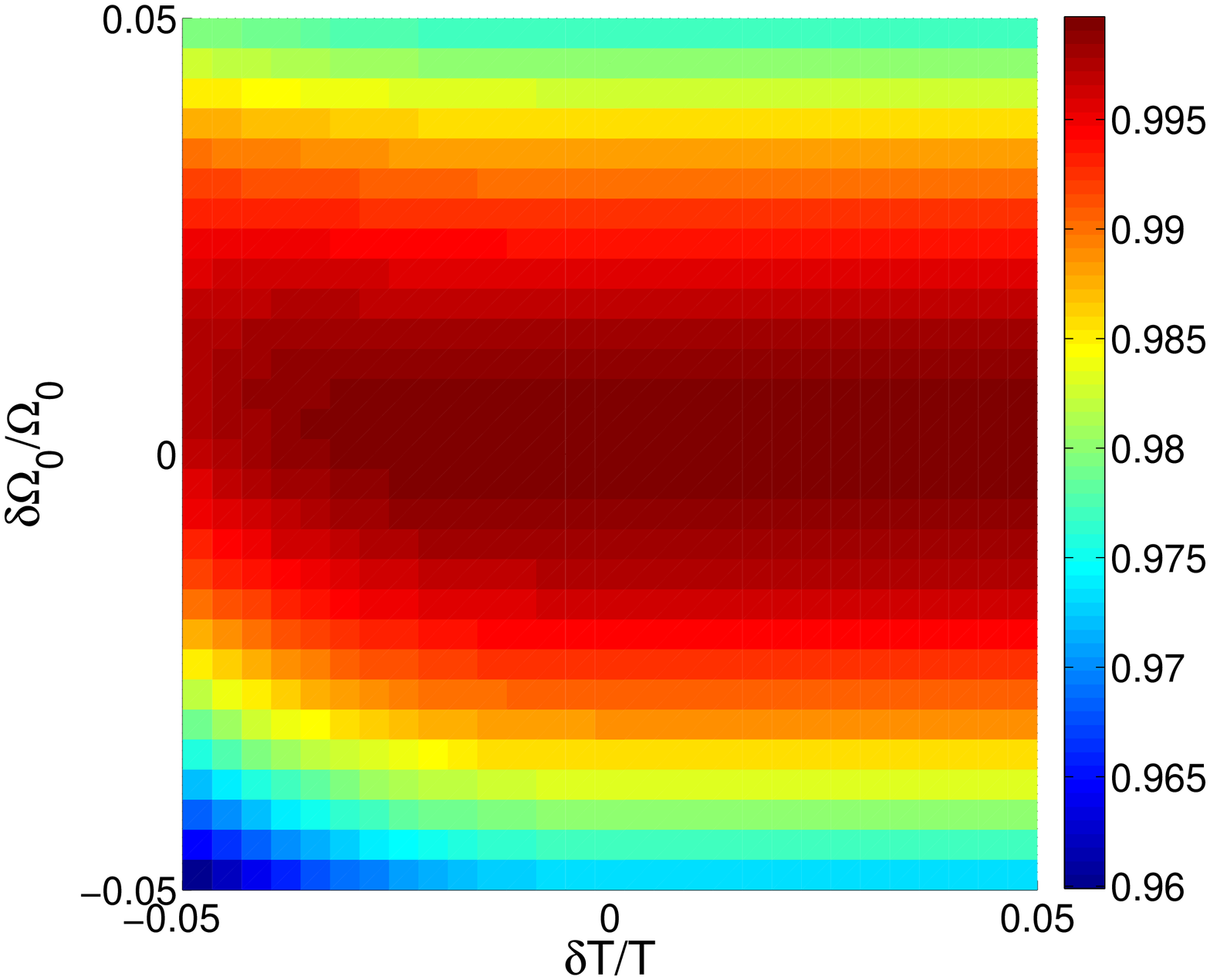}}
 \caption{
    The fidelity $F_{1}$ of the target state $-|\psi_{1}\rangle$ versus the variations of
    (a) $\lambda$ and $v$;
    (b) $\Omega_{0}$ and $T$.
          }
 \label{Fdeltax}
\end{figure}

The robustness against operational imperfection is also a main factor for the feasibility of the scheme because most of the
parameters are hard to faultlessly achieve in experiment. Hence, we calculate the fidelity by varying error parameters of the mismatch
among the laser amplitude, the interaction time, and coupling constants. And we define $\delta x=x'-x$ as the deviation of parameter $x$.
The fidelity versus the variations in different parameters are shown in Fig. \ref{Fdeltax}.
As shown in the figures, the scheme is robust against the variations of $\lambda$, $v$, and $T$ ($T=t_{f}$ is the total
operation time). But it is a little sensitive to the variation of laser amplitude $\Omega_{0}$
(a deviation $\delta \Omega_{0}/\Omega_{0}=\pm5\%$ causes a reduction about $3\%$ in the fidelity).
That is because the laser amplitude is strongly related to $\epsilon$ whose deviation greatly influence the target state's fidelity as shown in Fig. \ref{FE}.
Fortunately, that is not a serious problem to realize the scheme because it is not a challenge to
accurately control the laser amplitude in experiment.

In a real experiment, the cesium atoms are applicable to implement the
schemes. And an almost perfect fiber-cavity
coupling with an efficiency larger than $99.9\%$ can be realized
using fiber-taper coupling to high-Q silica microspheres
\cite{SMSTJKOJPKJVPrl03}. In case of that the fiber loss at $852$-nm wavelength is
only about $2.2$ dB/Km \cite{KJGVFPDTGSBIeee04}, the
fiber decay rate is only $0.152$ MHz. While in recent experimental
conditions \cite{JRBHJKPra03,SMSTJKKJVKWGEWHJKPra05}, it is
predicted that a strong atom-cavity coupling
$\lambda=2\pi\times750$ MHz is achieved. That means the fiber decay can actually
be neglected in a real experiment with these parameters. And choosing another set of parameters
$(\lambda,\Gamma,\kappa)=(2500,10,10)$ MHz in the microsphere
cavity QED which is reported in
\cite{SMSTJKKJVKWGEWHJKPra05,MJHFGSLBMBPNat06}, the fidelity of
the $\pi$ phase gate is higher than $99\%$.

\section{Conclusion}
``Shortcuts to adiabatic passage'' has shown its charm in a wide range of applications in quantum information processing
and quantum computation. It successfully breaks the time limit in an adiabatic process by providing alternative fast and
robust processes which reproduce the same final populations, or even the same final state as the adiabatic process in a finite and
shorter time. As we mentioned above, constructing STAP usually has
to deal with the quantum phase transition during the evolution, while how to make use of the phase transition
for quantum information processing and computation in cavity QED systems still has not been involved.
In this paper, we studied theoretically applications of STAP for the implementation of
$\pi$ phase gates and the creation of cluster states in distant-atom
systems which have advantages in long-distant quantum information processing and quantum computation. Though the interaction time required
for the gate operation in the scheme is limited by the Zeno condition,
it is still shorter than that in similar schemes via adiabatic passage and quantum Zeno dynamics.
Synthesizes each kind of influence factors of a theoretical scheme's realizability, the present
shortcut scheme might be a more reliable choice in experiment, because the scheme is not only fast,
but also robust against fluctuations of the parameters and decoherence caused by atomic spontaneous
emission and photon loss. In response to the interesting phenomenon that in the adiabatic dark state evolution,
the oscillation number is closely related to a special parameter $\zeta$. We have offered analysis and
discussion, in as much detail as possible.
The paper is promising and
might enlighten applications of shortcuts to adiabaticity for quantum phase gates in cavity QED
and even other experimental systems.

\section*{ACKNOWLEDGEMENT}

  This work was supported by the National Natural Science Foundation of China under Grants No.
11105030 and No. 11374054, the Foundation of Ministry of Education
of China under Grant No. 212085, and the Major State Basic Research
Development Program of China under Grant No. 2012CB921601.

\newpage

\end{document}